\documentstyle[epsf]{article}
\begin{document}

\renewcommand{\baselinestretch}{1.5}
\large

\begin{center}
{\Large {\bf A QED Shower   
Including the Next-to-leading Logarithm
 Correction
in $e^+e^-$ Annihilation} }
\vskip 0.8in
   T. MUNEHISA
\vskip 0.2in
Faculty of Engineering, Yamanashi University

Takeda, Kofu, Yamanashi 400-8511, Japan
\vskip 0.3in
       J. FUJIMOTO, Y.KURIHARA and  Y. SHIMIZU
\vskip 0.2in
National Laboratory for High Energy Physics(KEK)
 
Oho 1-1 Tsukuba, Ibaraki 305-0801, Japan
\vskip 0.5in
\end{center}
\vskip 0.5in
{\bf ABSTRACT}

 We develop an event generator, NLL-QEDPS, based on
the  QED shower  including the next-to-leading
logarithm correction in the  $e^+e^-$ annihilation.
The shower model  is  the   Monte Carlo technique to  solve
 the renormalization group equation so that
 they can calculate contributions of $\alpha^m\log^n(S/m_e^2)$ for
any $m$ and  $n$ systematically. Here $\alpha$ is the  QED coupling,
$m_e$ is the  mass of electron and $S$ is the  square of the 
 total energy in the $e^+e^-$ system.
While the previous  QEDPS   is
limited to the leading logarithm approximation
which includes only contributions of $(\alpha\log(S/m_e^2))^n$,
 the  model developed here contains terms of  $\alpha(\alpha\log(S/m_e^2))^n$,
 the the next-to-leading logarithm correction.
 The shower model is formulated
  for the  initial radiation in the $e^+e^-$ annihilation.
The  generator based on it gives us events with $q^2$, which is a virtual
mass squared of the virtual photon and/or Z-boson,
in accuracy of $0.04\%$, except for small $q^2/S$.
\eject

\def\GeV{\hbox{GeV}}
\def\MeV{\hbox{MeV}}

\noindent{\bf 1  \ \ Introduction }

In high energy reactions with electron beams, 
 it is important to study
radiative corrections\cite{rc}.
For this study event generators are indispensable tools.
We  have made the event generator, QEDPS
\cite{ps1}-\cite{ps4},
 for radiative corrections
in the $e^+e^-$ annihilation based on the shower model,
 which can radiate any number 
of photons.
However,  this model is limited
to the leading logarithm(LL) approximation.
In this paper we develop a shower model
 in the next-to-leading logarithm(NLL)  approximation\cite{nll1}-\cite{nll4}.
 The magnitude of the NLL order correction is of $\alpha^2/\pi^2\log(S/m_e^2)$,
which is about  0.0001, if $\sqrt{S} $ is $100\GeV$.
 So the NLL shower    might be irrelevant to actual
measurements.  However this  is   wrong
since   contributions due to soft photons are large.
They are estimated to be
  $\alpha^2/\pi^2\log(S/m_e^2)\log^2(E_{\gamma}/\sqrt{S})$,
which is about 0.005 if  the measured energy $E_{\gamma}$ for
a observed photon is $100\MeV$.
 This value  is not negligible
in   precise experiments.

In this paper we limit the event generator in  the NLL approximation
to the $e^+e^-$ annihilation, especially to the radiative process on the
initial state.
Applications of the NLL shower to other process such as the Bhabha scattering
\cite{ps2} are  discussed in other papers.

Our study is completely based on the renormalization group equation(RGE),
which has been developed well in QCD\cite{qcd}.
First we  clarify meanings of the NLL order approximation.
Let us consider  a dimensionless observable $F(Q^2/\mu^2,\alpha_0)$
with the mass scale $Q^2$ and the renormalization point $\mu^2$.     
If the coupling  $\alpha_0$  is  small  at $\mu^2$
 and the  ratio $Q^2/\mu^2$ is large,
 the RGE shows us that  $F(Q^2/\mu^2,\alpha_0)$ can be expanded by
the coupling constant after summing terms of $[\alpha_0 \log(Q^2/\mu^2)]^n $
for all $n$.
\begin{eqnarray}
 F(Q^2/\mu^2,\alpha_0)&=&  F^{(1)}(\alpha_0\log(Q^2/\mu^2))+
\alpha_0 F^{(2)}(\alpha_0\log(Q^2/\mu^2))
\nonumber \\
&+& \alpha_0^2F^{(3)}(\alpha_0\log(Q^2/\mu^2))
+ ... 
\end{eqnarray}
The first term is the LL order approximation.
If the the second term is included, the approximation is of
the NLL order. 

Sect.2 contains  discussions on  the RGE and
the formulas  needed for later sections. 
In Sect.3 we  apply them 
to the $e^+e^-$ annihilation process and  give  quantities  such as
the anomalous dimensions explicitly.
In Sect.4 we  formulate the shower model briefly.
In Sect.5 we  discuss the singular behavior of the NLL order
correction. The  simple perturbative expansion  breaks down due to effects
of soft photons
so that more sophisticated techniques are introduced. 
Then the effective shower model  in the LL order is 
given. At this stage the model has the same form as that in the LL order model.
 But some constraints are imposed    so that
it has included some contributions of the NLL correction.
Sect.6 contains results of the effective shower model.
In Sect.7 we present the explicit form of the second order $P$ function used in
the NLL shower model.
In Sect.8  we construct the event generator by
defining   kinematical variables  in terms of the variable  in
the shower model.
Also some problems on the construction
 are pointed out. 
The NLL approximation needs the second order coefficient  in
 the $\beta-$function of the coupling, but
we   drops contributions from this coefficient, which 
 is   discussed in Sect.9.
In Sect.10 we  present a method to compensate  results by 
the shower  because they  contain the $Q^2$-independent contribution due to 
the constraint.
Sect.11  is devoted to conclusions and discussions, where
numerical results in our study are summarized as well as limitations of our
model are given. Also we make some comments on applications of our model to
QCD. 

We present three appendices for some technical parts of our model.
Appendix A gives us   relations between the usual perturbative expansion
and  the RGE. In Appendix B we discuss the approximation that is 
made in order to get analytical expressions for results by the shower model.
In Appendix C we  present compact descriptions for
the shower algorithm, which we apply to compensating results for
 the $Q^2$-independent contribution.

\vskip 2cm
\noindent{\bf 2  \ \ Renormalization group equation}

In  the RGE\cite{qcd}, the value of the coupling depends on
 $\mu^2$, so that it is not constant, but the 
 function
of  $\mu^2$, $\alpha(\mu^2)$. Then the derivative of the coupling
 by $\mu^2$ is given by the  function of the coupling only.
$$ \mu^2 \frac{d\alpha(\mu^2)}{d \mu^2} = \beta(\alpha(\mu^2)). $$
Solving this equation, we obtain that
\begin{equation}
 \log(Q^2/\mu^2) = \int^{\bar\alpha}_{\alpha_0}
 \frac{1}{\beta(\alpha)} d\alpha ,
 \label{beta1}
\end{equation}
$$  \bar\alpha =\alpha(Q^2),\ \ \ \alpha_0=\alpha(\mu^2).
$$
Here $ \bar\alpha $ is called  the running coupling.
If one applies the RGE to the dimensionless observable
 $ F(Q^2/\mu^2,\alpha_0)$, 
the following equation is  obtained.
\begin{equation}
(\mu^2 \partial /\partial \mu^2 +\beta(\alpha_0)\partial /\partial\alpha_0
 -\gamma(\alpha_0) )  F(Q^2/\mu^2,\alpha_0) =0.
\end{equation}
 $ \gamma(\alpha)$ is the  anomalous dimension which is  the
 function of the coupling $\alpha$ only and
 depends on a process.
Then we solve this equation to  obtain that
\begin{eqnarray}
F(Q^2/\mu^2,\alpha_0)
&=&\exp(-\int^{\bar\alpha}_{\alpha_0}\gamma(\alpha)/\beta(\alpha) d\alpha)
F(1,\bar\alpha).
\label{rge0}
\end{eqnarray}
In the NLL order, 
$$ \beta(\alpha)= \beta_1\alpha^2 +\beta_2\alpha^3 , $$
$$ \beta_1= \frac{1}{3\pi}, \ \ \ \beta_2=\frac{1}{2\pi^2},
$$
$$ \gamma(\alpha)= \gamma_1\alpha +\gamma_2\alpha^2 . $$
Solving Eq.(\ref{beta1}) on the running coupling,
\begin{equation}
 \beta_1\log(Q^2/\mu^2) = -(\frac{1}{\bar\alpha}
-\frac{1}{\alpha_0})
 -\frac{\beta_2}{\beta_1 } \log( \bar\alpha /\alpha_0).
\end{equation}
If $\alpha_0 $ is  small but  $\alpha_0\log(Q^2/\mu^2) $ is large,
we  keep  any term of  $[\alpha_0\log(Q^2/\mu^2)]^n $  and
drop terms of   $\alpha_0^K\log(Q^2/\mu^2)(K \geq 3) $.
Then we obtain the explicit formula for the  coupling at $Q^2$
in the NLL order.
\begin{equation}
 \bar\alpha=\frac{\alpha_0}{1-\alpha_0\beta_1\log(Q^2/\mu^2) }
\lbrace 1- \frac{\alpha_0\beta_2}{\beta_1}
 \frac{\log\lbrack 1-\alpha_0\beta_1\log(Q^2/\mu^2)\rbrack
}{1-\alpha_0\beta_1\log(Q^2/\mu^2) }
 \rbrace .
\label{alpha1}
\end{equation}

  The integral inside  the exponential of (\ref{rge0})
 on the anomalous dimension   is  carried out  in the NLL order
\begin{equation}
I_{\gamma}= \int^{\bar\alpha}_{\alpha_0} \gamma(\alpha)/\beta(\alpha) d\alpha
 = \frac{\gamma_1}{\beta_1}\log(\bar\alpha/\alpha_0)
+(\frac{\gamma_2}{\beta_1} -\frac{\gamma_1\beta_2}{\beta_1^2})
(\bar\alpha-\alpha_0).
\label{rge1}
\end{equation}
The $Q^2$ dependence of  $F(Q^2/\mu^2,\alpha_0)$ can be
expressed by  $\beta_1,\beta_2, \gamma_1$, $\gamma_2 $ and
 notations used in Appendix A.
\begin{eqnarray}
F(Q^2/\mu^2,\alpha_0)&=& \exp\lbrack -\int_{\alpha_0}^{\overline{\alpha}}
d\alpha \frac{\gamma(\alpha)}{\beta(\alpha)}\rbrack F(1,\overline{\alpha})
\nonumber \\
&=& \exp\lbrack -\frac{\gamma_1}{\beta_1}
\log(\frac{\overline{\alpha}}{\alpha_0})
 -(\frac{\gamma_2}{\beta_1} -\frac{\gamma_1\beta_2}{\beta_1^2})
(\overline{\alpha}-\alpha_0) \rbrack
(f_0 + f_1^0\overline{\alpha}).
\end{eqnarray}
To  obtain the formula for the  $Q^2$ dependence,
 we take the  ratio of $F(Q^2/\mu^2,\alpha_0)$
and $F(1,\alpha_0)$.
\begin{eqnarray}
& &F(Q^2/\mu^2,\alpha_0)/F(1,\alpha_0)
\nonumber \\
&=& \exp\lbrack -\frac{\gamma_1}{\beta_1}
\log(\frac{\overline{\alpha}}{\alpha_0})
 -(\frac{\gamma_2}{\beta_1} -\frac{\gamma_1\beta_2}{\beta_1^2})
(\overline{\alpha}-\alpha_0) \rbrack
(f_0 + f_1^0\overline{\alpha})/ (f_0 + f_1^0\alpha_0).
\end{eqnarray}
Here since $(f_0 + f_1^0\overline{\alpha})/ (f_0 + f_1^0\alpha_0)$ can be
approximated by $1+\frac{f_1^0}{f_0}(\overline{\alpha}-\alpha_0)$,
\begin{eqnarray}
& &F(Q^2/\mu^2,\alpha_0)/F(1,\alpha_0)
\nonumber \\
&=& \exp\lbrack -\frac{\gamma_1}{\beta_1}
\log(\frac{\overline{\alpha}}{\alpha_0})
 -(\frac{\gamma_2}{\beta_1} -\frac{\gamma_1\beta_2}{\beta_1^2})
(\overline{\alpha}-\alpha_0) \rbrack
\lbrack 1 + \frac{f_1^0}{f_0}(\overline{\alpha}-\alpha_0) \rbrack .
\end{eqnarray}
Then we  expand terms of $\overline{\alpha}-\alpha_0$ in the
exponent and drop terms of $(\overline{\alpha}-\alpha_0)^K$($K \ge 2$).
\begin{eqnarray}
& &F(Q^2/\mu^2,\alpha_0)/F(1,\alpha_0)
\nonumber \\
&=& \exp\lbrack -\frac{\gamma_1}{\beta_1}
\log(\frac{\overline{\alpha}}{\alpha_0}) \rbrack
\lbrace 1 +
 (-\frac{\gamma_2}{\beta_1} +\frac{\gamma_1\beta_2}{\beta_1^2}
+\frac{f_1^0}{f_0})
(\overline{\alpha}-\alpha_0) \rbrace .
\label{rge2}
\end{eqnarray}
Finally  we  describe
 the explicit expression of the
$Q^2$-dependence for $ F(Q^2/\mu^2,\alpha_0)$.
\begin{eqnarray}
& &F(Q^2/\mu^2,\alpha_0)/F(1,\alpha_0)
\nonumber \\
&=& \exp\lbrack \frac{\gamma_1}{\beta_1}
\log(1-\alpha_0\beta_1\log(Q^2/\mu^2))
+\frac{\gamma_1\beta_2\alpha_0}{\beta_1^2}
\frac{\log(1-\alpha_0\beta_1\log(Q^2/\mu^2))}
{1-\alpha_0\beta_1\log(Q^2/\mu^2)}
  \rbrack
\nonumber \\
& \lbrace &1 +
 (-\frac{\gamma_2}{\beta_1} +\frac{\gamma_1\beta_2}{\beta_1^2}
+\frac{f_1^0}{f_0})
\frac{\beta_1\alpha_0^2\log(Q^2/\mu^2)}
{1-\alpha_0\beta_1\log(Q^2/\mu^2) }
\rbrace.
\label{rge3}
\end{eqnarray}
The above  is the  fundamental equation for the NLL shower model.

\vskip 3cm
\noindent{\bf 3 \ \  Annihilation }

 In this section we present explicit formulas for
 the annihilation cross section.
When $\sigma_0(Q^2)$ is the  bare cross section, i.e. the cross section
without any radiative correction, 
the  observed cross section is expressed by
the structure function $D(x,Q^2)$ and the coefficient function 
$C(x,\alpha)$ \cite{dy1}.
\begin{eqnarray}
\frac{d\sigma_{obs}(S,Q^2)}{dQ^2}=\sigma_0(Q^2)\frac{1}{S}
 \int^1_0 dx_1\frac{1}{x_1} \int^1_0 dx_2\frac{1}{x_2}
 D(x_1,Q^2)D(x_2,Q^2) C(z,\bar\alpha),
\label{crss1}
\end{eqnarray}
$$ z=\frac{\tau}{x_1x_2}= \frac{Q^2}{s},
\ \ \ \  \tau =\frac{Q^2}{S}, \ \  s=x_1x_2 S,$$
where $S$ is the total energy squared.

 In oder to  solve  the RGE in the NLL order analytically, we   take
 moments of  Eq.(\ref{crss1}).
\begin{eqnarray}
\frac{d\sigma_{obs}(n,Q^2)}{dQ^2}= \frac{\sigma_0(Q^2)}{Q^2}
 D(n,Q^2)D(n,Q^2) C(n,\bar\alpha).
\end{eqnarray}
Here they are given by taking moments.
\begin{eqnarray}
 \frac{d\sigma_{obs}(n)}{dQ^2}= \int_0^1 d\tau 
 \frac{d\sigma_{obs}(S,Q^2)}{dQ^2}\tau^n,
\end{eqnarray}
\begin{eqnarray}
 D(n,Q^2) = \int_0^1 \frac{dx}{x}x^n D(x,Q^2) ,
\end{eqnarray}
\begin{eqnarray}
 C(n,\alpha) = \int_0^1 \frac{dz}{z}z^nC(z,\alpha).
\end{eqnarray}
$D(n,Q^2) $ corresponds to the exponential term  in Eq.(\ref{rge0}), while
$C(1,\bar\alpha)$ does to $F(1,\bar\alpha)$ so that 
\begin{eqnarray}
 D(n,Q^2)= \exp[ -\int_{\alpha_0}^{\bar\alpha}
\frac{\gamma(n,\alpha)}{\beta(\alpha)} d \alpha ],
\label{Dsol1}
\end{eqnarray}
\begin{eqnarray}
\gamma(n,\alpha)=\gamma_1(n)\alpha+\gamma_2(n)\alpha^2,
\end{eqnarray}
\begin{eqnarray}
 C(n,\alpha)= 1+\frac{\alpha}{2\pi}C_1(n).
\end{eqnarray}
Then we obtain  the  $Q^2$-dependence of the cross section, 
as in Eq.(\ref{rge2}).
\begin{eqnarray}
& &   \frac{1}{\sigma_0(Q^2)} Q^2\frac{d\sigma_{obs}(n,Q^2)}{dQ^2} 
= \exp\lbrack -2\frac{\gamma_1(n)}{\beta_1}
\log(\frac{\overline{\alpha}}{\alpha_0}) \rbrack
\nonumber \\
& \times &
\lbrace 1 +
 (-2\frac{\gamma_2(n)}{\beta_1} +2\frac{\gamma_1(n)\beta_2}{\beta_1^2}
+\frac{1}{2\pi}C_1(n))
(\overline{\alpha}-\alpha_0) \rbrace .
\end{eqnarray}

 Here we summarize $\gamma_1(n), \gamma_2(n) $ and other quantities 
 needed for the annihilation process.
In order to get the   DGLAP equation, which is used in the shower model,
we  replace the variable $\mu^2$ by $Q^2$ in the RGE, so that
the equation of $D(n,Q^2)$ on   $Q^2$ is
 \begin{equation}
 Q^2\frac{d }{d Q^2} D(n,Q^2) =
   (Q^2 \frac{\partial  }{\partial  Q^2}+
   \beta(\bar\alpha)\frac{ \partial }{\partial \bar\alpha })D(n,Q^2)
= -\gamma(n,\bar\alpha) D(n,Q^2).
\label{rge5}
\end{equation}
Of course   $D(n,Q^2)$ of Eq.(\ref{Dsol1}) satisfies the above equation.
The inverse Mellin transformation of the above
is called the DGLAP equation.
\begin{eqnarray}
Q^2 \frac{d D(x,Q^2)}{d Q^2} = \int_x^1 \frac{dy}{y} P(x/y,\bar\alpha))
D(y,Q^2).
\end{eqnarray}
Here   $P(x,\alpha)$ is called  $P$ function, which is
 defined by the anomalous dimension
$ \gamma(n,\alpha)$;
$$  \gamma(n,\alpha) = -\int_0^1 dx x^{n-1} P(x,\alpha).
$$
$P(x,\alpha)  $ is divided into the LL and the NLL terms.
\begin{eqnarray}
 P(x,\alpha) = \frac{\alpha}{2\pi}P^{(1)}(x)
 + (\frac{\alpha}{2\pi})^2P^{(2)}(x) ,
\end{eqnarray}
$$  P^{(1)}(x) =  P_+(x),
$$
$$  P(x) = \frac{1+x^2}{1-x}.
$$
 Here we introduce the $+$notation for a function $f(x)$.
$$  f_+(x) = f(x) - \delta(1-x) \int_0^1 f(y) dy . $$
The explicit form of $\gamma_1(n)$ is useful for examining  
the model.
$$ \gamma_1(n) = \frac{-1}{2\pi}\int_0^1 dx x^{n-1} P^{(1)}(x)
=\frac{1}{2\pi}(2S_1(n-1) -\frac{3}{2}
+\frac{1}{n}+\frac{1}{n+1}).
$$
Here we introduce functions on the summation.
$$ S_m(n) = \sum_{k=1}^n \frac{1}{k^m}  \ \ \ \ {\rm for} \ \  n \ge 1 , $$
$$ S_m(n) =0 \ \ \ \ {\rm for} \ \  n \le 0 . $$
 The NLL term in the moment expression
is  not so compact so that we  present only
$P^{(2)}(x)$, which is found in Refs.\cite{nll1},\cite{nll4}\footnote{
We drop terms of branching of $e^-$ into $e^+$, which
are quite small.}.

$$ \gamma_2(n) = \frac{-1}{(2\pi)^2}\int_0^1 dx x^{n-1} P^{(2)}(x),
$$
$$  P^{(2)}(x) = P_{a+}(x) + P_{b+}(x),
$$
\begin{eqnarray}
  P(x)_a &=& -P(x)\log(x)\log(1-x)-(\frac{3}{1-x}+2x)\log(x)
 \nonumber \\
& - &
\frac{1}{2}(1+x)\log^2(x)-5(1-x),
\end{eqnarray}
\begin{eqnarray}
   P_b(x) = \frac{2}{3}[ P(x)(-\log(x)-\frac{5}{3})-2(1-x) ],
\end{eqnarray}
$$   C(x,\alpha)=\delta(1-x) + \frac{\alpha}{2\pi} C_{1}(x) . $$
Here $C_1(x)$ has been calculated for the Drell-Yan process in QCD\cite{dy1},
which is the sum of $C_{ep+}$ in the deep-inelastic scattering
 and an additional term $ \Delta C(x)$.
$$ C_{1}(x) = 2C_{ep+}(x) + \Delta C(x),
$$
\begin{eqnarray}
C_{ep}(x) = P(x)[ \log(\frac{1-x}{x})-\frac{3}{4} ] +\frac{9+5x}{4},
\end{eqnarray}
\begin{eqnarray}
  \Delta(x) = 2P(x)\log(1-x)-\frac{3}{1-x}-6-4x,
\end{eqnarray}
\begin{eqnarray}
  \Delta C(x) = \Delta_+(x)+\delta(1-x)(
-\frac{7}{2}+\frac{4\pi^2}{3}).
\end{eqnarray}

In our model we make use of  the scheme dependence,
which says that only the combination of 
$
 -2\frac{\gamma_2(n)}{\beta_1}+\frac{1}{2\pi}C_1(n)
$
can be predicted by the RGE, but each quantity is not.
By making use of this freedom,
we can put $\tilde{C}_1(n) =0$.
\begin{eqnarray}
  -2\frac{\gamma_2(n)}{\beta_1}+\frac{1}{2\pi}C_1(n)
= -2\frac{\tilde{\gamma}_2(n)}{\beta_1}+\frac{1}{2\pi}\tilde{C}_1(n)
= -2\frac{\tilde{\gamma}_2(n)}{\beta_1}.
\end{eqnarray}
That is
\begin{eqnarray}
\tilde{\gamma}_2(n)=
  \gamma_2(n)-\frac{\beta_1}{4\pi}C_1(n).
\end{eqnarray}
Or 
\begin{eqnarray}
\tilde{P}^{(2)}(x)= P^{(2)}(x)+ \pi\beta_1C_1(x).
\label{e32}
\end{eqnarray}
In our model we  use $\tilde{P}^{(2)}(x)$, which means that
the hard cross section is not employed.
\vskip 3cm
\noindent
{\bf 4 \ \ Shower model}

 The shower model that we call in this paper is
the  Monte Carlo method  to
solve the DGLAP equation by
 repeating the branching that the electron branches out into the child electron
and the photon, where  a variable $x$ and
 a virtual mass squared $K^2$
 are generated. Here  $x$ is the energy fraction of the child to one of the 
parent, while $K^2$ is the absolute value of the virtual mass squared of
the child.
 So the moment distribution of $x_b^{n-1}=(x_1x_2 \cdot \cdot \cdot x_L)^{n-1}$ 
 calculated by
the shower model  agrees with the analytic result
of the RGE completely within the statistical error.
Here $x_i$ is $x$ at  the $i-$th branching and $L$ is a number of 
branchings in one branching process.

The shower model needs the scheme to cutoff the
infrared singularity, though it is arbitrary.
Since we would like to apply the shower model to
 the event generator that  produces
electrons, photons and other particles in simulations,
we adopt the following cutoff scheme.
$$  x < 1-\mu^2/K^2  . $$
The definition of $x$, which  is  necessary to construct the generator,
 is given  in Sect.8.

Also our shower model employs the double cascade scheme, in order that
the electron and the positron make the branching process
 independently\cite{nll2},\cite{ps1},\cite{ps3}. In this scheme we impose the 
constraint to $x$.
$$  1-x > K^2/Q^2 .$$

 By using these constraints, we can apply the Monte Carlo method to 
generate $x$ and $K^2$, as described in Ref.\cite{ps1}.
\vskip 3cm

\noindent {\bf 5 \ \  The singular behavior of the NLL order correction}

In Sect.3 we presented the  $P$ function in  Eq.(\ref{e32}),
 which we use in the shower model.
As $x \rightarrow 1$, the most  singular behavior of this function is
$$ (\frac{\alpha(K^2)}{2\pi} )^2 \ 8\pi  \beta_1  \  \frac{\log(1-x)}{1-x}.
$$
 This is dangerous, since it can become
larger than the singular LL term, which is  
$$ \frac{\alpha(K^2)}{2\pi}  \frac{2}{1-x}.
$$
  In order to make the branching stable,  it was
suggested  to use the running coupling $\alpha((1-x)K^2) $ instead of
$\alpha(K^2) $ in order to
include term of $ \alpha^n\frac{\log^n(1-x)}{1-x}$ into the model
in Ref.\cite{nll1}. By this replacement, the singular behavior of $P$ function
is
$$  \frac{\alpha(K^2)}{2\pi} \frac{2}{1-x}
 +A (\frac{\alpha(K^2)}{2\pi})^2\frac{\log(1-x)}{1-x}$$
$$ = \frac{\alpha_0}{2\pi[1-\beta_1\alpha_0\log(K^2/\mu^2)]}
\frac{2}{1-x} +A(\frac{\alpha(K^2)}{2\pi})^2\frac{\log(1-x)}{1-x}$$
$$ = \frac{\alpha_0}{2\pi[1-\beta_1\alpha_0\log((1-x)K^2/\mu^2)]}
\frac{2}{1-x} +(A-4\pi\beta_1) (\frac{\alpha(K^2)}{2\pi})^2
\frac{\log(1-x)}{1-x}
$$
$$
+ O(\alpha_0^3) .
$$
In the deep inelastic scattering, indeed $A=4\pi\beta_1$, so that
the dangerous term disappears after the replacement.
On the other hand we have $A=8\pi\beta_1$ in the annihilation.
By this method  $4\pi\beta_1$ of them can be
included into the running coupling.
 We have to include  the remnant $4\pi\beta_1$ terms into the effective
LL form in order to remove  the dangerous term in the second order
$P$ function.
 This can be done by taking account of the kinematical constraint in
the annihilation as follows.

In the annihilation,
both electron and positron to radiate photons.
$$e^-(P_1)+e^+(P_2)\rightarrow e^-(p_1)+e^+(p_2)+X
\rightarrow \gamma(q) +X    . $$
This implies that the spacelike virtual electron($p_1$) and
positron ($p_2$) annihilate into the virtual photon($q$). 
 Although we can calculate $q^2$  by  $p_1$, $p_2$, as seen in Fig.1,
 we make the approximation  that
\begin{equation}
  q^2\approx x_{b1}x_{b2}(1-t_1)(1-t_2)S. 
\label{q2a}
\end{equation}
Here
$t_1=-p_1^2/Q^2(t_2=-p_2^2/Q^2)$.
The accuracy  of this approximation is discussed in Sect.8.

Therefore our shower model calculates 
 moments with respect to the variable $x_b(1-t)$ .
   The $n-$th moment $(x_b(1-t))^{n-1}$  
should agree  with 
 the structure function $D_s(n,Q^2)$ on
the electron(See ref.\cite{ps3}).
\begin{eqnarray}
&&D_s(n,Q^2)=\Pi(Q^2,\mu^2)+ \int_0^1 {dt \over t} (1-t)^{n-1} \Pi(Q^2,tQ^2)
                       \nonumber\\
&&\qquad\qquad\times
\int_0^1 dx {\alpha((1-x)tQ^2) \over 2 \pi }
 P(x)x^{n-1}\theta(1-x - t)\theta((1-x)t-\epsilon)
                                          \nonumber \\
&&\times
  \exp\lbrace  \int^{t}_{0} { d t' \over t' }
\int dx' {\alpha((1-x')t'Q^2) \over 2 \pi }  P(x')( x'^{n-1} - 1)
\nonumber \\
&&\times
  \theta (1-x' - t')\theta((1-x')t'-\epsilon)
\rbrace ,  
\label{ds0}
\end{eqnarray}
\begin{eqnarray}
\Pi(Q^2,Q_0^2)&=&
\exp[ -\int^{1}_{Q_0^2/Q^2}
    { d t \over t } \int dx {\alpha((1-x)tQ^2) \over 2 \pi }
\nonumber \\
&&\times P(x)\theta(1-x-t)\theta((1-x)t-\epsilon)] .
\label{ds1}
\end{eqnarray}
\noindent
Here $\epsilon=\mu^2/Q^2$. $\Pi(Q^2,Q^2_0)$ is the non-branching probability
that the electron does not branch  for possible virtual mass squared
between $Q^2$ and $ Q^2_0$.
In Eq.(\ref{ds0}) the first term represents the no-branching case so that
 the moment is unity for any $n$.
The front term on the exponential  does
the last branching, while  the exponential  appears
after repeating branchings.
The reason for the special form on the last branching is that
there the virtual mass squared is involved in
the moment, as seen in Eq.(\ref{q2a}).

In order to obtain the expression that is possible to be 
 calculated analytically,
we approximate  $ (1-t)^n $ by $\theta(1/n -t) $.
\begin{eqnarray}
D_s(n,Q^2) &=&
\exp\lbrace \int^{1}_{0}
{ d t \over t } \int_0^{1} dx {\alpha((1-x)tQ^2) \over 2 \pi }
\nonumber \\
&& \times  P(x)[ (x(1-t))^{n-1} - 1]
    \theta(1-x - t)\theta((1-x)t-\epsilon) \rbrace .
\label{ds2}
\end{eqnarray}
\noindent
The error due to this approximation is  discussed in Appendix B.
 We write  $D_s(n,Q^2)= \exp\lbrack I_s(n,Q^2) \rbrack $, and 
we  perform the integrals for $I_s(n,Q^2)$.
 \begin{eqnarray}
&&I_s(n,Q^2) = 
\nonumber \\
& = &\int_0^1 dz (z^{n-1}-1)  \int^{1}_{0}
{ d t \over t } \int_0^1 dx  \delta( z-x(1-t))
{\alpha((1-x)tQ^2) \over 2 \pi }
 P(x) \nonumber \\
& \times & \lbrack (x(1-t))^{n-1} - 1 \rbrack
    \theta(1-x - t)\theta((1-x)t-\epsilon) .
\end{eqnarray}
\noindent
Then we integrate $I_s(n,Q^2)$ over $x$ and $ t$, where
we assume that $n$  is not so large so that contributions from regions of
$z \sim 1 $ can be neglected.
 Also we neglect $O(\epsilon)$ terms.   
Finally we obtain 
\begin{eqnarray}
& & I_s(n,Q^2)
\nonumber \\
&=&\frac{1}{2\pi}
\int_0^1 dx(x^{n-1}-1) P(x)\frac{1}{\beta_1}
\log(\frac{\overline{\alpha}}{\alpha_0})
+\frac{\overline{\alpha}}{2\pi}\int_0^1 dx(x^{n-1}-1)\Delta P(x),
\label{p1d}
\end{eqnarray}
where
\begin{eqnarray}
& & \Delta P(x) = 2P(x)\log(1-x)-\frac{1}{1-x}\log{x}
\nonumber \\
&+&(1+x)\log(1+\sqrt{x})
+\frac{1}{2}(1+x)\log{x} -\sqrt{x}+x .
\end{eqnarray}
In Eq.(\ref{p1d}), the first term is the LL order while
the the second order is the NLL order, though it contains
the $Q^2$-independent contribution.
By noting that  $\overline{\alpha}=(\overline{\alpha}-\alpha_0)+\alpha_0$
and
\begin{equation}
  \int_{\mu^2/Q^2}^1 \frac{dt }{t} [ \alpha(tQ^2) ]^2
= \frac{1}{\beta_1} (\overline{\alpha} -\alpha_0 ),
\end{equation}
one can see that the second term has $(\alpha/2\pi)^2 \log(1-x)/(1-x) $
and its coefficient is $8\pi\beta_1$ as expected.

\vskip 3cm

\noindent{\bf  6  \ \ Results by the shower in the effective LL order  }

Summarizing the discussion in the previous section,
Eq.(\ref{p1d}) was derived by adopting the three schemes:

\noindent
1) to use $(1-x)K^2$ for the argument of  the  running coupling, i.e.
at the branching  we employ the following coupling,
$$ \alpha((1-x)K^2)=\frac{\alpha_0}{1-\beta_1\alpha_0\log((1-x)K^2/\mu^2)}.
$$
As pointed out in Ref.\cite{nll1}, $(1-x)K^2$ is about the transverse momentum
squared at the branching.
\vskip 0.5cm
\noindent
2)  the double cascade scheme\cite{nll2}, where we impose the
constraint of 
$$   (1-x) > K^2/Q^2 . $$
\vskip 0.5cm
\noindent
3) to define   $q^2$ in the annihilation as
$$  q^2 = x_{b1}(1-t_1) x_{b2}(1-t_2)S,
$$
$$x_b=1\times x_1 \times \cdot \cdot \cdot \times x_L.
$$
\vskip 0.5cm

The moment of $x_b(1-t)$ by  this shower in Eq.(\ref{p1d})  is  calculated in 
the analytic form.
\begin{eqnarray}
& &I_s(n,Q^2) 
= \frac{1}{2\pi \beta_1}\log(\frac{\overline{\alpha}}{\alpha_0})
\lbrack -2S_1(n-1)-\frac{1}{n}-\frac{1}{n+1}+\frac{3}{2} \rbrack
\nonumber \\
&+& \frac{\overline{\alpha}}{2\pi}\lbrack
2(S_1^2(n-1)+S_2(n-1)+\frac{1}{n}S_1(n)+\frac{1}{n+1}S_1(n+1)
-\frac{7}{4} )
\nonumber \\
&-& S_2(n-1)+\frac{1}{n}(-S_1(n)+S_1(2n))
+\frac{1}{n+1}(-S_1(n+1)+S_1(2n+2))
\nonumber \\
&-&\frac{1}{2}(\frac{1}{n^2}+\frac{1}{(n+1)^2})
-\frac{2}{2n+1}+\frac{1}{n+1} \rbrack .
\nonumber \\
\label{eq:llm}
\end{eqnarray}
Some comparisons between analytic calculations, Eq.(\ref{eq:llm}), 
 and results by the shower model
 are shown  in  Table 1.
There we assumed that $\mu^2=0.25\times 10^{-6}\GeV^2$ and
$\alpha_0=1/137$.
In  Monte Carlo simulations a total number of events is $10^{8}$
and errors are estimated by calculating the variance in
10 data sets of 
 $10^7$ events.
The agreement of order of $10^{-5}$ justifies 
our discussion.

The structure function given by Eq.(\ref{eq:llm}) 
has the $Q^2$-independent  contribution, $D^f(n)$, which does not
vanish  at $Q^2=\mu^2$
because
the  term  of $\overline{\alpha}$ in Eq.(\ref{eq:llm})  remains then.
 $D^f(n)$ is  obtained by replacing $\overline{\alpha}$
 by $\alpha_0$  in Eq.(\ref{p1d}).
\begin{eqnarray}
& & D^f(n) =\exp\lbrace 
\frac{\alpha_0}{2\pi}\int_0^1 dx(x^{n-1}-1)\Delta P(x) \rbrace .
\label{fin1}
\end{eqnarray}
In order to calculate the absolute value as well as the $Q^2$-dependence
for the structure function by the shower model,
 we have to compensate it for  the contribution due to
$D^f(n)$.
The method for the  compensation   is discussed in Sect.10.

\vskip 3cm
\noindent{\bf  7 \ \  Effective  $P^{(2)}(x)$}

In this section we give the second order  $P$ function 
 $P^{(2)eff}(x)$, which is used in 
the shower model.  Since our model imposes that
 the coefficient function  is zero 
so that the NLL contribution can be
given  by  $\tilde{P}^{(2)}(x)$ in Eq.(\ref{e32}).
As discussed in Sect.5 and 6, our shower model contains  the NLL
contribution, $\Delta P(x)$, 
 through $\alpha((1-x)K^2)$, the double cascade scheme and
the definition of $q^2$,  as described in Eq.(\ref{p1d}).
In the NLL shower we employ these schemes so that the structure function
$ D(n,Q^2)= exp[  I(n,Q^2) ]$ is given by
\begin{eqnarray}
& &  I(n,Q^2) =
\int_0^1 dx \int_0^1 \frac{dt}{t}
\theta(1-x-t)\theta((1-x)t-\epsilon)
\lbrack (1-t)^{n-1}x^{n-1} -1 \rbrack
\nonumber \\
&&\lbrace \frac{\alpha(t(1-x)Q^2)}{2\pi}
P(x) +
\frac{\alpha(t(1-x)Q^2)^2}{(2\pi)^2}P^{(2)eff}(x) \rbrace  .
\label{nll2}
\end{eqnarray}
By performing the integral similar to one in the previous section,
 we have
\begin{eqnarray}
&&  I(n,Q^2) = \frac{1}{2\pi \beta_1}
 \log(\frac{\overline{\alpha}}{\alpha_0})
 \int_0^1 dx (x^{n-1} -1 )P(x)
\nonumber \\
&&
+ \frac{\overline{\alpha}}{2\pi} \int_0^1 (x^{n-1} -1 )\Delta P(x) 
  + \frac{1}{(2\pi)^2 \beta_1}
 (\overline{\alpha}-\alpha_0)
 \int_0^1 dx (x^{n-1} -1 )P^{(2)eff}(x) 
\nonumber
\\
&=&    \frac{1}{2\pi \beta_1}
 \log(\frac{\overline{\alpha}}{\alpha_0})
 \int_0^1 dx (x^{n-1} -1 )P(x)
\nonumber \\
&&
  + \frac{1}{(2\pi)^2 \beta_1}
 (\overline{\alpha}-\alpha_0)
 \int_0^1 dx (x^{n-1} -1 )\lbrack P^{(2)eff}(x) 
+ 2\pi\beta_1 \Delta P(x) \rbrack 
\nonumber \\
&&
+ \frac{\alpha_0}{2\pi} \int_0^1 dx (x^{n-1} -1 )\Delta P(x) .
\label{nll4}
\end{eqnarray}
Here we neglected terms of order of $\alpha_0^2$ or $\overline{\alpha}^2$.
$\tilde{P}^{(2)}(x)$ in Eq.(\ref{e32}) should equal to
$P^{(2)eff}(x) + 2\pi\beta_1 \Delta P(x)$ .
Therefore   $P^{(2)eff}(x)$ is given by
\begin{eqnarray}
& &P^{(2)eff}(x) = \tilde{P}^{(2)}(x) - 2\pi\beta_1\Delta P(x)
\nonumber \\
&=& P_a(x)+P_b(x)
\nonumber \\
&+&2\pi\beta_1
\lbrack -\frac{3}{1-x} 
-P(x)\log(x)+\frac{\log(x)}{1-x}
-(1+x)\log(x+\sqrt{x})+\sqrt{x}-x\rbrack,
\label{eq:p2net}
\end{eqnarray}
which is free from the singular term 
$\log(1-x)/(1-x)$, as is  expected.

\vskip 3cm

\noindent{\bf  8 \ \ Event generator }

 In this section we present the event generator based on the shower model,
 which was described in Sect.4-7.
 In order to determine four momenta of the produced particles
 we 
 define  $x$ to be a fraction of $+$($-$) component of
 lightcone variables for electrons(positron).
\begin{eqnarray}
  x &=& \frac{p_+ }{P_+}(  \frac{p_-}{P_-}) ,
 \\
  p_{\pm} &=& \frac{ E\pm p_z}{\sqrt{2}}.
\end{eqnarray}
Here four momenta are denoted as $(p_x,p_y,p_z,E) $ and
$P_-(P_+)$ denotes a lightcone variable of the initial electron(positron)
\cite{ps1}.
At the branching of  $ e^-(y,-K^2) \rightarrow e^-(xy, -K'^2) +
\gamma(y(1-x),0) $, the momentum conservation imposes the following
equation.
\begin{equation}
   -K^2 =\frac{-K'^2}{x}+ \frac{\vec{k_T}^2}{x(1-x)},
\end{equation}
where $\vec{k_T}$ is the transverse momentum, $(p_x,p_y)$.
Here note that
the electron during the branching process  is spacelike. 
Our cutoff scheme, $ x < 1-\mu^2/K'^2$,  equals to 
$\vec{k_T}^2 \ge \mu^2$, if $K^2 \ll K'^2$.

 Using  an arbitrary azimuthal angle $\phi$, 
 $K^2, K'^2$ and $x$ one can  determine four
momenta of the electron and  the  photon after the branching.
\begin{eqnarray}
   p_{e'\mu}&=&(xP_+,\frac{-K'^2+\vec{k}_T^2}{x}, \vec{k_T} ) ,
\\
   p_{\gamma \mu}&=&((1-x)P_+,\frac{\vec{k}_T^2}{1-x}, -\vec{k_T} ) ,
\\
  k_{x} &=& k_T \cos \phi, \ \ \ k_{y} = k_T \sin \phi  ,
\\
  k_{T}^2 &=& (1-x)(-xK^2 + K'^2 ) .
\end{eqnarray}
These equations determine four momenta of all particles completely.
This implies that in the annihilation process  the four momentum $q$
of the virtual photon and/or  Z-boson is the sum of 
  momenta $p_1$ and $p_2$ of the electron and positron after
the branching process, 
that is $q=p_1+p_2$(See Fig.1).
Then  the virtual mass squared of the four momentum is
\begin{eqnarray}
 q^2 &=& p_1^2 + p_2^2 +2p_1p_2 
\nonumber \\
 &=& -K_1^2 - K_2^2 +2(p_{1+}p_{2-}+p_{1-}p_{2+}-\vec{p}_{1T}\vec{p}_{2T})
\nonumber \\
 &=& -K_1^2 - K_2^2 +2(x_{b1}P_{1+}x_{b2}P_{2-}
 +\frac{-K_1^2+\vec{p}_{1T}^2}{2x_{b1}P_{1+}}
 \frac{-K_2^2+\vec{p}_{2T}^2}{2x_{b2}P_{2-}}
 -2\vec{p}_{1T}\vec{p}_{2T})
\nonumber \\
  &=&-K_1^2-K_2^2 +x_{b1}x_{b2} S 
 +\frac{(-K_1^2+\vec{p}_{1T}^2) (-K_2^2+\vec{p}_{2T}^2)}{x_{b1}x_{b2}S}
-2\vec{p}_{1T}\vec{p}_{2T} .
\end{eqnarray}
The variable $\tau=q^2/S$ equals
\begin{eqnarray}
 \tau &=& -t_1-t_2 +x_{b1}x_{b2} 
 +\frac{(K_1^2+\vec{p}_{1T}^2) (K_2^2+\vec{p}_{2T}^2)}{x_{b1}x_{b2}S^2}
-2\vec{p}_{1T}\vec{p}_{2T} .
\label{tau1}
\end{eqnarray}
Here we used $t= K^2/S$.
In the generator the ratio $\tau$  is given by
the above equation, but not by $x_{b1}x_{b2}$.
Although  the RGE  predicts  moments  on $\tau$, as described 
in Sect.3,
$\tau $ of (\ref{tau1}) is not a good variable for  the shower model,
because it gives us moments on $\tau'=x_{b_1}(1-t_1)x_{b2}(1-t_2)$,
as discussed in Sect.5. We present the detailed discussion on the
accuracy of the generator, which uses $\tau$.

Let us discuss differences between moments on $\tau$ and $\tau'$.
First note that the last term, $2\vec{p}_{1T}\vec{p}_{2T}$,
 in the above equation  is  zero if 
averages are took,  because angles between these
vectors are arbitrary.
Next  $\tau$ can be negative while $\tau'$ is always positive.
Of course the negative $\tau$ is unphysical so that the event with
the negative   $\tau$ is abandoned.
We  introduce a variable $\overline{\tau}$ that
can be negative and whose moments are possible to be
calculated analytically, which is
\begin{equation}
   \overline{\tau} = (x_{b1}-t_1)(x_{b2}-t_2).
\end{equation}
In this definition $\overline{\tau} $
 might be negative. A case of being negative
is counted as an event, but these negative values are replaced by
zero.
 Also the generator 
 can fail to make four momenta for 
the virtual photon and/or Z-boson, i.e. $\tau$ is negative.
 The failed case  is counted as
an event and $\tau $ is set to be zero. 
Results on  moments of $\tau$, $\overline{\tau}$ and $\tau'$
 for a total number of $10^8$ are presented  in Table 2.
Differences on data for $\tau $ in the generator and $\overline{\tau}$
in the  shower are quite small and less than $ 10^{-5}$.
So we can conclude that moments of $ \tau $ in the generator is accounted by
those of $\overline{\tau}$.
 We can estimate analytically 
differences of moments 
 on  $\overline{\tau}$ and $\tau'$, which  are of order of $10^{-3}$ and
decrease rapidly  as $n$ rises. They are
\begin{eqnarray}
&& D_d(n,Q^2) = 
\exp\lbrace 2 \frac{1}{2\pi}
\int_0^1 \frac{dt}{t}\int_0^1 dx \theta(t(1-x)-\epsilon)
\theta(1-x-t)
 \nonumber \\
&& 
\alpha(t(1-x)Q^2)P(x)\lbrack (1-t)^{n-1} x^{n-1}-1 \rbrack
\rbrace
 \nonumber \\
&& 
 -  \exp\lbrace 
  2 \frac{1}{2\pi}
\int_0^1 \frac{dt}{t}\int_0^1 dx \theta(t(1-x)-\epsilon)
\theta(1-x-t)
\nonumber \\
&&
\alpha(t(1-x)Q^2)P(x)\lbrack \theta(x-t)(x-t)^{n-1}-1 \rbrack 
\rbrace .
\end{eqnarray}
Here for simplicity we neglect  $P^{(2)eff}(x)$.
If we neglect the running effect on $\alpha$ and approximate
$D_d(n,Q^2)$  by the difference between  the first terms
 in expansions by $\alpha$,
\begin{eqnarray}
& &D_d(n,Q^2)\approx \frac{2\alpha_0}{2\pi}\int_0^1 \frac{dt}{t} \int_0^1 dx
 P(x)
\theta(t(1-x)-\epsilon)\theta((1-x)-t)
\nonumber \\
&& \hskip 3cm \times [ x^{n-1}(1-t)^{n-1} - \theta(x-t)(x-t)^{n-1} ].
\label{Dd}
\end{eqnarray}
Under these approximations,  $D_d(n,Q^2)$ is independent of
$Q^2$.
The values of  the expression (\ref{Dd})  presented in  Table 3
can account for the difference on the moment of $\tau'$ and $\overline{\tau}$.
For an example, the moment of $n=2$
 in Table 2 is  $0.93212$, and $0.93118$, while
 Table 3 shows the difference $0.00093$, which agrees with the difference.

\vskip 3cm
\noindent{\bf 9 \ \   $\beta$-function }

In this section we  estimate contributions by 
the second order correction  $\beta_2$ in  the  beta function. 
Using   Eq.(\ref{alpha1})
 for the running coupling, $ \overline{\alpha}-\alpha_0 $ is given by
\begin{equation}
 \overline{\alpha}-\alpha_0
=\frac{\beta_1\alpha_0^2\log(Q^2\mu^2)}
{1-\alpha_0\beta_1\log(Q^2/\mu^2) }.
\end{equation}
Inserting the explicit expression of the running coupling,
the integral on the anomalous dimension  $I_\gamma$  in Eq.(\ref{rge1})
 becomes
\begin{eqnarray}
&&I_\gamma=
  \frac{\gamma_1}{\beta_1}\log
 \frac{1}{1-\alpha_0\beta_1\log(Q^2/\mu^2) }
\nonumber \\
 &-& \frac{\alpha_0\beta_2\gamma_1}{\beta_1^2}
 \frac{\log\lbrack 1-\alpha_0\beta_1\log(Q^2/\mu^2) \rbrack }
 {1-\alpha_0\beta_1\log(Q^2/\mu^2) }
+(\frac{\gamma_2}{\beta_1} -\frac{\gamma_1\beta_2}{\beta_1^2})
\frac{\beta_1\alpha_0^2\log(Q^2\mu^2)}
{1-\alpha_0\beta_1\log(Q^2/\mu^2) }.
\end{eqnarray}
In the QED process $\alpha_0\beta_1\log(Q^2/\mu^2) $
is small so that the logarithm  of the second term is
approximated as follows.
\begin{equation}
 \log\lbrack 1-\alpha_0\beta_1\log(Q^2/\mu^2) \rbrack 
\approx
 -\alpha_0\beta_1\log(Q^2/\mu^2) .
\end{equation}
This  leads to  cancellation of the terms with
$\beta_2$.
\begin{equation}
I_\gamma \approx
  \frac{\gamma_1}{\beta_1}\log
 \frac{1}{1-\alpha_0\beta_1\log(Q^2/\mu^2) }
+\gamma_2\alpha_0^2
\frac{\log(Q^2\mu^2)}
{1-\alpha_0\beta_1\log(Q^2/\mu^2) } .
\end{equation}
The leading  term in the neglected terms is
$$
\frac{\gamma_1\beta_2\alpha_0^3}{2}
\frac{\log(Q^2/\mu^2)}{1-\alpha_0\beta_1\log(Q^2/\mu^2) },
$$
which
 is less than  $10^{-6}$ if one uses actual values for
 $ \alpha_0, \beta_1,\beta_2$,
$\gamma_1$ and $Q^2/\mu^2=$ $(100\GeV/0.5\MeV)^2=$ $2.5\times  10^{11}$.
Therefore we can neglect the term with $\beta_2$ safely.

\vskip 3cm
\noindent{\bf 10 \ \   $Q^2$-independent contribution}

 As discussed in Sect.5,  the shower model 
contains  the  $Q^2$-independent contribution. In other words, the structure
function is not $\delta(1-x)$ at $Q^2=\mu^2$, but its moment is given
by $D^f(n) $ of Eq.(\ref{fin1}).
In study of the radiative corrections on QED, the absolute value of the
cross section, not its $Q^2$-dependence, has to be calculated.
So  we would like to  
compensate the structure function for
  the the $Q^2$-independent contribution $D^f(n)$.
On the moment we only divide results by  $D^{f}(n)$. 
$$ D^{cmp}(n,Q^2) = D(n,Q^2)/D^{f}(n).$$
Here $ D^{cmp}(x,Q^2) $ is the structure function after the
compensation. 
However, in the generator we need  the inverse Mellin transformation.
The product of moments is
equivalent to the convolution integral of the function in the transformation
so 
\begin{eqnarray}
& & D^{cmp}(x,Q^2) = \int_x^1 \frac{dy}{y} D(y,Q^2) D^{\overline{f}}(x/y),
\nonumber \\
& &(D^f(n))^{-1} = \int_0^1 dx x^{n-1} D^{\overline{f}}(x).
\label{eq:conv2}
\end{eqnarray}
In the shower the convolution integral is realized by the procedure
that $x$ of the initial electron is fixed according to
the probability $D^{\overline{f}}(x)$ and then we make the branching
process, which induces the structure function.

For performing  this  we are confronted with  three problems. First
the explicit form of $D^{\overline{f}}(x)$ is difficult to calculate.
Here we use the shower algorithm to get $x$, which  is  described 
 in detail in Appendix C.

Second the function  $D^{\overline{f}}(x)$ is not positive necessarily so that
it could be used as the probability.
Equivalently the splitting  function $-\Delta P(x)$ in
$$
( D^{f}(n))^{-1}=\exp[ -\frac{\alpha_0}{2\pi}
\int_0^1 dx( x^{n-1}-1)\Delta P(x) ],
$$
is negative at  $x$ near zero.
Since   $-\Delta P(x)$ is concentrated
near $x \sim 1$,
 we must be  contented with  an approximated $\Delta P^{A}(x)$, which
is modified near $x \sim 0$. The approximated $\Delta P^{A}(x)$ 
is given by
\begin{eqnarray}
\Delta  P^{A}(x) = \theta(x - x_c) \Delta  P(x) . 
\end{eqnarray}
Here $x_c$ is fixed to satisfy the condition,
$$   \int_0^{x_c} dy  \Delta  P(y) = 0. $$

The third problem is that the total energy squared $S$  of $e^+e^-$ system
is changed by
$ x_1^f x_2^f S $, where $ x_1^f$ and $ x_2^f $ are $x-$fractions of the
initial electron and positron after the compensation by $D^{\overline{f}}(x)$.
Since errors due to this problem are very small numerically,
we neglect this problem.

Next we  examine  numerical results related to  $D^f(n)$. 
First we  compare $(D^f(n))^{-1}$ with
\begin{eqnarray}
D^{f,A}(n)=\exp[-\frac{\alpha_0}{2\pi} \int_0^1 (x^{n-1} -1)\Delta P^{A}(x) ],
\label{dfa1}
\end{eqnarray}
 in Table 4.
One finds  the agreement between results by the shower algorithm
 and
the moment $(D^f(n))^{-1}$ except those for small $n$.
The differences are less than $0.01 \%$.
In order to confirm that these differences are due to the second
problem, we calculate the moment  $D^{f,A}(n)$ through
the numerical integration to show these results in the same table.
The agreement between values of the second and the third column
supports our discussion strongly.

\vskip 3cm
\noindent{\bf 11 \ \  Conclusions and discussions }

In this paper we have formulated the shower model including the NLL correction
in the $e^+e^-$ annihilation and developed the generator.
 Results on  the $Q^2$-dependence of  moments  by  the  generator 
are summarized
in Tables 5, 6 and 7.
Here the moment for $n=1$ is fixed to be unity because
it is the event  normalization,  so that the analytic value is 
the ratio between the $n-$th  moment and the first moment.
  Analytic results in  the NLL order  
  are      $0.03 \% $ of the LL order ones for  small  $n$, but
 they increase as $n$ rises and are about
$0.2\%$ for $n = 100$.
 The effect of the NLL order is small, but could not be neglected
in precise experiments as at LEP or in future colliders.
Results of our shower model agree with the
analytic calculations of the NLL oder in accuracy of
$0.04 \%$.  
 The agreement in  the generator is worse for small $n$,
 because events might not satisfy with the kinematics 
constructed by the shower model. 
 Simply speaking, the constructed value
for $q^2$ becomes negative in these events.
 Table 7 shows differences between moments by the shower model
and the analytic one, and those between  moments by the
generator and the analytic one in the NLL order.
 Values in the  table  indicate the magnitude of the systematic error
 in the present shower and the generator in the NLL order.

Next we mention some comments on limitations of our model.
First
 the accuracy found in $q^2$ distributions may not be common to
 other distributions such as the transverse momentum distributions of
radiated photons.
One of  reasons is that we do not include the cross section for the emission 
 of a photon with a large transverse momentum.

 Also we neglect effects by the three-body decay in the shower
because there is a less interest on detailed distributions of photons.
Further we neglect the mixing $P^{(2)}(x)$,
which is a contribution that the electron radiates into
the spacelike positron with the pair creation, since
its effect is expected to be very small.

Third our shower model is limited to the non-singlet case where
there are no contributions by radiations with the spacelike
photon. A reason for this limitation is that in experiments
one can exclude events with the electron positron pair easily,
which correspond to pure singlet radiations, as well as 
 that  they are quite small.
Finally notice that we use $S$ for the mass scale of the RGE, but not
$q^2$.

In this study we have examined our model in the moment form.
But from a experimental view, 
 analyses in the $x$-space are desired, which will be discussed
in coming papers. 
Also we have to discuss the accuracy of our generator in detail by applying
it to  several realistic processes such as the muon pair production or
the $Z-$Higgs production.

Finally we would like to stress that our study on the NLL shower
is quite important for QCD, where the NLL shower has been developed.
Because there has been no study on the $Q^2$-independent contributions by
the shower algorithm in QCD.
Also precise discussions 
have led  us to  the deeper understanding of the  shower models.
 Therefore our  study
stimulates the interest on further developments of QCD showers.

\vskip 3cm
\noindent{\bf Acknowledgements}
  
 We would like to thank our colleagues of KEK working group
(Minami-Tateya) and in LAPP for their interests and discussions.
Especially we   appreciate valuable comments by Prof. Kato.
This work has been done under the collaboration between KEK and
LAPP supported by Monbusho, Japan(No.07044097) and CNRS/IN2P3, France.

\vfill
\eject

\noindent
{\bf Appendix A}

In this appendix we  explain   results of
the RGE in terms of   the perturbative expansion.
In the  perturbative  expansion by 
 the coupling $\alpha_0$ at $ \mu^2$, 
the dimensionless quantity $F(Q^2/\mu^2,\alpha_0) $
 is  calculated up to some order of $\alpha_0$.
\begin{equation}
 F(Q^2/\mu^2,\alpha_0) = f_0 + f_1(Q^2/\mu^2)\alpha_0 + 
f_2(Q^2/\mu^2)\alpha^2_0+\cdot \cdot \cdot \ \ .
\end{equation}
Here we assume that the mass scale is only $Q^2$.
In order to compare results of the RGE with those of
  the perturbative calculation,
the ratio $F(Q^2/\mu^2,\alpha_0)/F(1,\alpha_0)$ should be used, 
because  the RGE can calculate the $Q^2$-dependence only.
 The ratio  is given by
\begin{eqnarray}
F(Q^2/\mu^2,\alpha_0)/F(1,\alpha_0) = 1+[ (f_1(Q^2/\mu^2)- f_1(1))/f_0 ]\alpha_0 +
 \nonumber
 \\
  \lbrack    (f_2(Q^2/\mu^2)- f_2(1))/f_0+f_1(1)^2/f_0^2
 -f_1(Q^2/\mu^2)f_1(1)/f_0^2  \rbrack
  \alpha^2_0  +\cdot \cdot \cdot \ \ .
\end{eqnarray}
The physical quantity contains only $\log(Q^2/\mu^2)$
and the order of power of the logarithm is less than the order
of the coupling constant. Therefore
\begin{eqnarray}
 f_1(Q^2/\mu^2) =f_1^{0}+ f_1^{1} \log(Q^2/\mu^2),
  \\
 f_2(Q^2/\mu^2) =f_2^{0}+ f_2^{1}\log(Q^2/\mu^2) + f_2^{2}\log^2(Q^2/\mu^2) .
\end{eqnarray}
If the ratio $F(Q^2/\mu^2,\alpha_0)/F(1,\alpha_0)$ is expressed 
in terms of
$ f_1^{0}, f_1^{1},f_2^{0}, f_2^{1}$ and $f_2^{2} $,
\begin{eqnarray}
& & F(Q^2/\mu^2,\alpha_0)/F(1,\alpha_0)= 
1+\alpha_0 \frac{f_1^{1}}{f_0}\log(Q^2/\mu^2)
\nonumber \\
&+&\alpha^2_0 \lbrace \frac{f_2^1}{f_0 }
\log(Q^2/\mu^2)
+\frac{f_2^2}{f_0}\log^2(Q^2/\mu^2)
-\frac{f_1^0 f_1^1}{(f_0)^2}\log(Q^2/\mu^2) 
\rbrace  .
\label{per1}
\end{eqnarray}
The result by the RGE has been discussed in Sect.3 and is
summarized by Eq.(\ref{rge3}).
If we drop terms of $\alpha_0^n$ ($ n \ge 3$) in this equation,
\begin{eqnarray}
& &F(Q^2/\mu^2,\alpha_0)/F(1,\alpha_0)
\nonumber \\
&\approx & 1-\alpha_0 \gamma_1 \log(Q^2/\mu^2)
+ \alpha_0^2 \log^2(Q^2/\mu^2)[ \frac{1}{2}(\gamma_1)^2
 -\gamma_1\beta_1 ]
\nonumber \\
&+& \alpha_0^2 \log(Q^2/\mu^2)[-\gamma_2+\frac{f_1^0\beta_1}{f_0} ] .
\label{rge4}
\end{eqnarray}
Since  the perturbative expansion (\ref{per1}) should agree with
 (\ref{rge4}) by the RGE, we obtain relations between $\gamma, \beta$ and
$f$.
$$   \frac{f_1^1}{f_0} = -\gamma_1 ,
$$
$$  \frac{f_2^2}{f_0} = \frac{1}{2}(\gamma_1)^2 -\gamma_1\beta_1 ,
$$
$$  \frac{f_2^1}{f_0}-\frac{f_1^0f_1^1}{(f_0)^2}
    = -\gamma_2 + \frac{f_1^0\beta_1}{f_0} .
$$

If we neglect the running effect i.e. $\beta_1=0$,  we have the simple relation
for $f_1^{1}$ and $ f_2^{2}$.
\begin{equation}
  f_2^{2}= \frac{(f_1^{1})^2}{2f_0}.
\end{equation}

\vskip 3cm
\noindent{\bf Appendix B}

In this appendix we  discuss  the analytic expression for
the moment of $x_b(1-t) $ in  the shower model.
We need some approximations in order to
get the analytic expression for the moment,
which  is defined in
 Eq.(\ref{ds0}) in Sect.5. 
The approximated, but analytic expression is given by
 the expression (\ref{ds2}), from which 
 we obtain  Eq.(\ref{p1d}).
A conclusion in this appendix is that
  the difference between the expression (\ref{ds0}) and
(\ref{ds2})  is  of order of
$\alpha^2$ and of $n^0$, and
does not have the  $Q^2$-dependence  except one due to 
the running effect.

First note that the expression  (\ref{ds2}) equals to
\begin{eqnarray}
&&\Pi(Q^2,\mu^2)+ \Pi(Q^2,\mu^2)\int_0^1 dt \frac{d}{dt}
\exp\lbrace \int^{t}_{0}
{ d t' \over t' } \int_0^{1} dx {\alpha((1-x)t'Q^2) \over 2 \pi }
 P(x) 
\nonumber \\
 && (x(1-t'))^{n-1}
     \theta(1-x - t')\theta((1-x)t'-\epsilon) \rbrace
\nonumber \\
&=&
\Pi(Q^2,\mu^2)+ \Pi(Q^2,\mu^2)\int_0^1  
{dt \over t} \int_0^{1} dx {\alpha((1-x)tQ^2) \over 2 \pi }
 P(x) 
\nonumber \\
 && (x(1-t))^{n-1}
     \theta(1-x - t)\theta((1-x)t-\epsilon)
\exp\lbrace \int^{t}_{0}
{ d t' \over t' } \int_0^{1} dx' {\alpha((1-x')t'Q^2) \over 2 \pi }
 P(x') 
\nonumber \\
 && (x'(1-t'))^{n-1}
     \theta(1-x' - t')\theta((1-x')t'-\epsilon) \rbrace .
\end{eqnarray}
Since the factorization can apply to the non-branching 
probability, we have
$$ \Pi(Q^2,\mu^2)=\Pi(Q^2,tQ^2)
\Pi(tQ^2,\mu^2) . $$
Then the expression (\ref{ds2}) becomes
\begin{eqnarray}
&&
\Pi(Q^2,\mu^2)+ \int_0^1
{dt \over t}\Pi(Q^2,tQ^2)
 \int_0^{1} dx {\alpha((1-x)tQ^2) \over 2 \pi } P(x)
\nonumber \\
 && (x(1-t))^{n-1}
     \theta(1-x - t)\theta((1-x)t-\epsilon)
\exp\lbrace \int^{t}_{0}
{ d t' \over t' } \int_0^{1} dx' {\alpha((1-x')t'Q^2) \over 2 \pi }
 P(x')
\nonumber \\
 &&\lbrack  (x'(1-t'))^{n-1}-1 \rbrack
     \theta(1-x' - t')\theta((1-x')t'-\epsilon) \rbrace .
\label{ds2x}
\end{eqnarray}

We examine a difference $Diff=$ (\ref{ds0})-(\ref{ds2}).
\begin{eqnarray}
&&Diff
\nonumber \\
&=& \int_0^1
{dt \over t}\Pi(Q^2,tQ^2)
 \int_0^{1} dx {\alpha((1-x)tQ^2) \over 2 \pi } P(x)
  (x(1-t))^{n-1}
     \theta(1-x - t)\theta((1-x)t-\epsilon)
\nonumber \\
&& \lbrace \exp\lbrack \int^{t}_{0}
{ d t' \over t' } \int_0^{1} dx' {\alpha((1-x')t'Q^2) \over 2 \pi }
 P(x')
  ( x'^{n-1} -1)
  \theta(1-x' - t')\theta((1-x')t'-\epsilon) \rbrack
\nonumber \\
&-&
\exp\lbrack \int^{t}_{0}
{ d t' \over t' } \int_0^{1} dx' {\alpha((1-x')t'Q^2) \over 2 \pi }
 P(x')
  ( (x'(1-t'))^{n-1} -1)
\nonumber \\
&& \hskip 5cm  \theta(1-x' - t')\theta((1-x')t'-\epsilon) \rbrack \rbrace .
\end{eqnarray}
If the exponential is expanded by $\alpha$,
\begin{eqnarray}
&&Diff
\nonumber \\
&\approx & \int_0^1
{dt \over t}
 \int_0^{1} dx {\alpha((1-x)tQ^2) \over 2 \pi } P(x)
  (x(1-t))^{n-1}
     \theta(1-x - t)\theta((1-x)t-\epsilon)
\nonumber \\
&& \lbrack \int^{t}_{0}
{ d t' \over t' } \int_0^{1} dx' {\alpha((1-x')t'Q^2) \over 2 \pi }
 P(x')
   x'^{n-1}(1-(1-t')^{n-1}))
\nonumber \\
&& \hskip 5cm  \theta(1-x' - t')\theta((1-x')t'-\epsilon) \rbrack .
\label{ds3}
\end{eqnarray}
Here  we would like to  show  that
this difference is not proportional to $\log(\epsilon)$ 
and finite as $n$ increases.
But
it is difficult to obtain the analytic expression for  Eq.(\ref{ds3})
 so that we  calculate the simpler expression by
replacing the running coupling by  $\alpha_0$.
This approximation could not change the essential property of
Eq.(\ref{ds3}).
If this expression is finite as $\epsilon$ becomes zero,
it is not proportional to $\log(\epsilon)$.
  So we  
 set $\epsilon$   zero in Eq.(\ref{ds3}).
Then
the difference is approximated by
\begin{eqnarray}
&&Diff
\approx ({\alpha_0 \over 2 \pi })^2 \int_0^1
{dt \over t} \int_0^{1-t} dx  P(x) (x(1-t))^{n-1}
\nonumber \\
&&  \hskip 3cm \times  \int^{t}_{0} { d t' \over t' } \int_0^{1-t} dx' P(x')
 x'^{n-1}(1-(1-t')^{n-1}) .
\nonumber \\
\label{ds4}
\end{eqnarray}
This integral is possible to be expressed analytically,
which implies that Eq.(\ref{ds3}) is finite as  $\epsilon$ goes to  zero.
But this expression is too long to understand the property.
In order to confirm that Eq.(\ref{ds4})
 is finite as $n$
increases, we  present numerical values of them for various $n$
in Table 8, which
supports  our conclusions.
Summarizing  this appendix, the error by the approximation on  $D_s(n,Q^2)$
 is of order of $\alpha^2$ and $n^0$  so that
we can neglect this difference safely.

\vskip 3cm
\noindent{\bf Appendix C}

In this appendix we present the shower
algorithm to generate $x$ according to that
the moment is given by Eq.(\ref{dfa1}), which is
$$ exp[ \int_0^1 dx (x^{n-1}-1) I^f(x) ], \ \ \ 
   I^f(x) = -\frac{\alpha_0}{2\pi} \Delta P^A (x) . $$
We introduce $\overline{D}^f(n,x_{max})$, which
is 
\begin{eqnarray}
 \overline{D}^f(n,x_{max})= exp[  \int_0^{x_{max}} dx x^{n-1} I^f(x) ] .
\end{eqnarray}
Since  the exponential is expanded to be
the infinite series,
\begin{eqnarray}
&& \overline{D}^f(n,x_{max}) =\exp( \int_0^{x_{max}} dx x^{n-1} I^f(x) )
= \sum_{k=0}^{\infty} [ \int_0^{x_{max}} dx x^{n-1} I^f(x) ]^k/k !
\nonumber \\
&=& \sum_{k=0}^{\infty}  \int_0^{x_{max}} dx_1 x_1^{n-1} I^f(x_1)
 \int_0^{x_1} dx_2 x_2^{n-1} I^f(x_2) \cdot \cdot 
 \int_0^{x_{k-1}} dx_k x_k^{n-1} I(x_k) ,
\end{eqnarray}
 we can separate contributions of the
no-branching and the branching  with one or more particles.
\begin{eqnarray}
 \overline{D}^f(n,x_{max})
   = 1 + \int_0^{x_{max}}dx x^{n-1} I^f(x) \overline{D}^f(n,x) . 
\end{eqnarray}
  By this equation the probability for no-branching $ Pr_{\rm N }(x_{max})$ is
 that
\begin{eqnarray}
  Pr_{\rm N}(x_{max}) = 1/\overline{D}^f(1,x_{max})
               =\exp( - \int_0^{x_{max}} dx I^f(x) ) .
\label{C2}
\end{eqnarray}
While the probability $Pr_{\rm E}(x) dx$ for the first branching
at $[x, x+dx] ( x< x_{max} )$
is given by
\begin{eqnarray}
&& Pr_{\rm E}(x)dx =  dx I^f(x)\overline{D}^f(1,x)/ \overline{D}^f(1,x_{max})
\nonumber \\
&& =dx I^f(x)\exp( - \int_x^{x_{max}} dy I^f(y) ) .
\label{C3}
\end{eqnarray}
By replacing $x_{max} $ by $x$, we can repeat the use
of Eqs.(\ref{C2}) and (\ref{C3}).
Then we obtain the  following iterative algorithm for the shower.

\begin{itemize}
\item{step(1)} Set $ x_m= 1 $ and $x_b=1$.

\item{step(2)}  Calculate a probability of stop,
$ Pr_N(x_m)=\exp(-\int_0^{x_m} dy I^f(y) ) $.

\item step(3) Generate a uniform random number $\xi$($ 0< \xi < 1$).
If $\xi$ is less than the probability, go to step(5). 

\item step(4)  Calculate $x$ that satisfies
$$  \xi= \exp( -\int_0^x dy I^f(y))  .
$$
Replace $x_b$ by $x_b x $.
Then set $x_m=x$ and go back to step(2).

\item step(5)
 Finish generating  one event. Then calculate $x_b^{n-1}$ for various $n$ and
accumulate them.

\end{itemize}

\vfill
\eject

\normalsize
\begin{table}[h]
\begin{tabular}{|c|c|c|c|c|c|} \hline
  & \multicolumn{2}{c|}{$Q^2=10^4GeV^2$}  
 &   \multicolumn{2}{c|}{$Q^2=10^6GeV^2$} \\ \hline
mom & Analytic  & Showers & Analytic & Showers \\ \hline \hline
    2 & 0.96536 & 0.96539 $\pm$0.16E-05 & 0.95836 & 0.95833 $\pm$0.86E-06 \\  \hline
    3 & 0.94740 & 0.94745 $\pm$0.20E-05 & 0.93670 & 0.93669 $\pm$0.11E-05 \\  \hline
    4 & 0.93525 & 0.93531 $\pm$0.21E-05 & 0.92200 & 0.92201 $\pm$0.13E-05 \\  \hline
    5 & 0.92609 & 0.92617 $\pm$0.22E-05 & 0.91091 & 0.91093 $\pm$0.14E-05 \\  \hline
    6 & 0.91878 & 0.91886 $\pm$0.23E-05 & 0.90202 & 0.90205 $\pm$0.16E-05 \\  \hline
    7 & 0.91270 & 0.91279 $\pm$0.24E-05 & 0.89464 & 0.89466 $\pm$0.17E-05 \\  \hline
    8 & 0.90751 & 0.90761 $\pm$0.25E-05 & 0.88833 & 0.88836 $\pm$0.18E-05 \\  \hline
    9 & 0.90301 & 0.90311 $\pm$0.26E-05 & 0.88283 & 0.88287 $\pm$0.19E-05 \\  \hline
   10 & 0.89902 & 0.89913 $\pm$0.27E-05 & 0.87797 & 0.87801 $\pm$0.20E-05 \\  \hline
   11 & 0.89546 & 0.89558 $\pm$0.28E-05 & 0.87363 & 0.87366 $\pm$0.20E-05 \\  \hline
   12 & 0.89225 & 0.89237 $\pm$0.29E-05 & 0.86969 & 0.86973 $\pm$0.21E-05 \\  \hline
   13 & 0.88932 & 0.88944 $\pm$0.29E-05 & 0.86611 & 0.86615 $\pm$0.22E-05 \\  \hline
   14 & 0.88663 & 0.88676 $\pm$0.30E-05 & 0.86282 & 0.86286 $\pm$0.22E-05 \\  \hline
   15 & 0.88415 & 0.88429 $\pm$0.31E-05 & 0.85978 & 0.85982 $\pm$0.23E-05 \\  \hline
   16 & 0.88186 & 0.88199 $\pm$0.31E-05 & 0.85696 & 0.85700 $\pm$0.23E-05 \\  \hline
   17 & 0.87971 & 0.87985 $\pm$0.32E-05 & 0.85433 & 0.85437 $\pm$0.24E-05 \\  \hline
   18 & 0.87771 & 0.87785 $\pm$0.32E-05 & 0.85187 & 0.85191 $\pm$0.24E-05 \\  \hline
   19 & 0.87583 & 0.87597 $\pm$0.33E-05 & 0.84956 & 0.84959 $\pm$0.24E-05 \\  \hline
   20 & 0.87405 & 0.87420 $\pm$0.33E-05 & 0.84738 & 0.84741 $\pm$0.25E-05 \\  \hline
   30 & 0.86047 & 0.86065 $\pm$0.35E-05 & 0.83061 & 0.83063 $\pm$0.27E-05 \\  \hline
   40 & 0.85130 & 0.85150 $\pm$0.35E-05 & 0.81923 & 0.81925 $\pm$0.29E-05 \\  \hline
   50 & 0.84445 & 0.84467 $\pm$0.35E-05 & 0.81070 & 0.81071 $\pm$0.30E-05 \\  \hline
   60 & 0.83904 & 0.83926 $\pm$0.34E-05 & 0.80392 & 0.80392 $\pm$0.31E-05 \\  \hline
   70 & 0.83458 & 0.83481 $\pm$0.33E-05 & 0.79832 & 0.79832 $\pm$0.32E-05 \\  \hline
   80 & 0.83080 & 0.83105 $\pm$0.32E-05 & 0.79357 & 0.79356 $\pm$0.33E-05 \\  \hline
   90 & 0.82754 & 0.82780 $\pm$0.32E-05 & 0.78945 & 0.78944 $\pm$0.34E-05 \\  \hline
  100 & 0.82468 & 0.82494 $\pm$0.31E-05 & 0.78583 & 0.78582 $\pm$0.34E-05 \\  \hline
\end{tabular}
\caption{  Numerical results of analytic calculations and the shower
 for the effective LL order P-function. Analytic results are given by
Eq.(38).  
Here $\mu^2=0.25\times 10^{-6} \GeV$, and $\alpha_0=1/137$.
The same values for them are used in  coming  tables.
    }
\end{table}
\begin{table}[h]
\begin{tabular}{|c|c|c|c|} \hline
mom & $x_{b1}(1-t_1)x_{b2}(1-t_2)$  & $(x_{b1}-t_1)(x_{b2}-t_2) $
& $q^2/S$ \\ \hline 
    2 & 0.93212 $\pm$0.34E-04 & 0.93118 $\pm$0.34E-04 & 0.93112 $\pm$0.34E-04 \\  \hline
    3 & 0.89787 $\pm$0.48E-04 & 0.89725 $\pm$0.48E-04 & 0.89721 $\pm$0.48E-04 \\  \hline
    4 & 0.87504 $\pm$0.58E-04 & 0.87458 $\pm$0.58E-04 & 0.87455 $\pm$0.58E-04 \\  \hline
    5 & 0.85804 $\pm$0.66E-04 & 0.85767 $\pm$0.66E-04 & 0.85765 $\pm$0.66E-04 \\  \hline
    6 & 0.84457 $\pm$0.71E-04 & 0.84426 $\pm$0.71E-04 & 0.84424 $\pm$0.71E-04 \\  \hline
    7 & 0.83346 $\pm$0.75E-04 & 0.83320 $\pm$0.75E-04 & 0.83318 $\pm$0.75E-04 \\  \hline
    8 & 0.82405 $\pm$0.79E-04 & 0.82382 $\pm$0.79E-04 & 0.82381 $\pm$0.79E-04 \\  \hline
    9 & 0.81590 $\pm$0.81E-04 & 0.81570 $\pm$0.81E-04 & 0.81569 $\pm$0.81E-04 \\  \hline
   10 & 0.80874 $\pm$0.84E-04 & 0.80856 $\pm$0.84E-04 & 0.80855 $\pm$0.84E-04 \\  \hline
   11 & 0.80236 $\pm$0.85E-04 & 0.80220 $\pm$0.86E-04 & 0.80219 $\pm$0.86E-04 \\  \hline
   12 & 0.79663 $\pm$0.87E-04 & 0.79648 $\pm$0.87E-04 & 0.79647 $\pm$0.87E-04 \\  \hline
   13 & 0.79142 $\pm$0.88E-04 & 0.79128 $\pm$0.88E-04 & 0.79127 $\pm$0.88E-04 \\  \hline
   14 & 0.78666 $\pm$0.90E-04 & 0.78653 $\pm$0.90E-04 & 0.78652 $\pm$0.90E-04 \\  \hline
   15 & 0.78228 $\pm$0.91E-04 & 0.78216 $\pm$0.91E-04 & 0.78215 $\pm$0.91E-04 \\  \hline
   16 & 0.77822 $\pm$0.92E-04 & 0.77811 $\pm$0.92E-04 & 0.77811 $\pm$0.92E-04 \\  \hline
   17 & 0.77446 $\pm$0.92E-04 & 0.77435 $\pm$0.92E-04 & 0.77435 $\pm$0.92E-04 \\  \hline
   18 & 0.77094 $\pm$0.93E-04 & 0.77084 $\pm$0.93E-04 & 0.77084 $\pm$0.93E-04 \\  \hline
   19 & 0.76765 $\pm$0.94E-04 & 0.76755 $\pm$0.94E-04 & 0.76755 $\pm$0.94E-04 \\  \hline
   20 & 0.76455 $\pm$0.95E-04 & 0.76446 $\pm$0.95E-04 & 0.76446 $\pm$0.95E-04 \\  \hline
   30 & 0.74103 $\pm$0.99E-04 & 0.74097 $\pm$0.99E-04 & 0.74096 $\pm$0.99E-04 \\  \hline
   40 & 0.72535 $\pm$0.10E-03 & 0.72531 $\pm$0.10E-03 & 0.72531 $\pm$0.10E-03 \\  \hline
   50 & 0.71376 $\pm$0.11E-03 & 0.71373 $\pm$0.11E-03 & 0.71373 $\pm$0.11E-03 \\  \hline
   60 & 0.70466 $\pm$0.11E-03 & 0.70463 $\pm$0.11E-03 & 0.70462 $\pm$0.11E-03 \\  \hline
   70 & 0.69720 $\pm$0.11E-03 & 0.69718 $\pm$0.11E-03 & 0.69718 $\pm$0.11E-03 \\  \hline
   80 & 0.69092 $\pm$0.11E-03 & 0.69090 $\pm$0.11E-03 & 0.69090 $\pm$0.11E-03 \\  \hline
   90 & 0.68553 $\pm$0.11E-03 & 0.68551 $\pm$0.11E-03 & 0.68551 $\pm$0.11E-03 \\  \hline
  100 & 0.68080 $\pm$0.11E-03 & 0.68079 $\pm$0.11E-03 & 0.68078 $\pm$0.11E-03 \\  \hline
\end{tabular}
\caption{ Numerical results of the shower and the generator in the effective
LL order. Columns with $x_{b1}(1-t_1)x_{b2}(1-t_2)$ and
 $(x_{b1}-t_1)(x_{b2}-t_2)$ 
 are results of the shower for these  variables, while the third columns 
 are
results of the generator for $q^2/S$.
   }
\end{table}
%
%
\begin{table}[h]
\begin{tabular}{|c|c|c|} \hline
mom &   $ D_d(n,Q^2)/(\alpha_0/2\pi)$ &  $D_d(n,Q^2)$ 
 \\ \hline
    1 &  1.6882 & 0.1961E-02 \\  \hline
    2 &  0.8009 & 0.9305E-03 \\  \hline
    3 &  0.5477 & 0.6363E-03 \\  \hline
    4 &  0.4223 & 0.4906E-03 \\  \hline
    5 &  0.3455 & 0.4014E-03 \\  \hline
    6 &  0.2931 & 0.3405E-03 \\  \hline
    7 &  0.2548 & 0.2960E-03 \\  \hline
    8 &  0.2255 & 0.2620E-03 \\  \hline
    9 &  0.2024 & 0.2351E-03 \\  \hline
   10 &  0.1836 & 0.2132E-03 \\  \hline
   11 &  0.1680 & 0.1952E-03 \\  \hline
   12 &  0.1549 & 0.1799E-03 \\  \hline
   13 &  0.1437 & 0.1669E-03 \\  \hline
   14 &  0.1340 & 0.1556E-03 \\  \hline
   15 &  0.1255 & 0.1458E-03 \\  \hline
   16 &  0.1181 & 0.1372E-03 \\  \hline
   17 &  0.1115 & 0.1295E-03 \\  \hline
   18 &  0.1056 & 0.1226E-03 \\  \hline
   19 &  0.1003 & 0.1165E-03 \\  \hline
   20 &  0.0955 & 0.1109E-03 \\  \hline
   30 &  0.0646 & 0.7501E-04 \\  \hline
   40 &  0.0488 & 0.5670E-04 \\  \hline
   50 &  0.0392 & 0.4556E-04 \\  \hline
   60 &  0.0328 & 0.3809E-04 \\  \hline
   70 &  0.0282 & 0.3272E-04 \\  \hline
   80 &  0.0247 & 0.2869E-04 \\  \hline
   90 &  0.0220 & 0.2553E-04 \\  \hline
  100 &  0.0198 & 0.2300E-04 \\  \hline
\end{tabular}
\caption{  Numerical results on $D_d(n,Q^2)$  of Eq.(57)
 in Sect.8.
    They 
are shown in the third column, while values in the second one are
calculated by dividing them by $\alpha_0/2\pi$.
 While in other tables the moment starts from 2, results for moment 1
are included here.
    }
\end{table}
\begin{table}[h]
\begin{tabular}{|c|c|c|c|} \hline
mom & $D^f(n)^{-1}$  & $D^{f,A}(n) $
& $D^{f,A}(n)_{shower}$ \\ \hline
    2 & 0.99714 & 0.99701 & 0.99700$\pm$0.82E-05 \\  \hline
    3 & 0.99450 & 0.99443 & 0.99442$\pm$0.15E-04 \\  \hline
    4 & 0.99221 & 0.99217 & 0.99216$\pm$0.20E-04 \\  \hline
    5 & 0.99017 & 0.99015 & 0.99015$\pm$0.24E-04 \\  \hline
    6 & 0.98835 & 0.98834 & 0.98833$\pm$0.27E-04 \\  \hline
    7 & 0.98669 & 0.98668 & 0.98667$\pm$0.30E-04 \\  \hline
    8 & 0.98516 & 0.98516 & 0.98515$\pm$0.32E-04 \\  \hline
    9 & 0.98375 & 0.98375 & 0.98374$\pm$0.35E-04 \\  \hline
   10 & 0.98243 & 0.98244 & 0.98243$\pm$0.37E-04 \\  \hline
   11 & 0.98120 & 0.98120 & 0.98119$\pm$0.38E-04 \\  \hline
   12 & 0.98004 & 0.98004 & 0.98003$\pm$0.40E-04 \\  \hline
   13 & 0.97894 & 0.97894 & 0.97893$\pm$0.42E-04 \\  \hline
   14 & 0.97790 & 0.97790 & 0.97789$\pm$0.43E-04 \\  \hline
   15 & 0.97690 & 0.97691 & 0.97690$\pm$0.44E-04 \\  \hline
   16 & 0.97596 & 0.97596 & 0.97595$\pm$0.45E-04 \\  \hline
   17 & 0.97505 & 0.97506 & 0.97505$\pm$0.46E-04 \\  \hline
   18 & 0.97418 & 0.97419 & 0.97418$\pm$0.48E-04 \\  \hline
   19 & 0.97334 & 0.97335 & 0.97334$\pm$0.49E-04 \\  \hline
   20 & 0.97254 & 0.97254 & 0.97253$\pm$0.49E-04 \\  \hline
   30 & 0.96577 & 0.96578 & 0.96577$\pm$0.57E-04 \\  \hline
   40 & 0.96054 & 0.96055 & 0.96054$\pm$0.62E-04 \\  \hline
   50 & 0.95624 & 0.95626 & 0.95625$\pm$0.66E-04 \\  \hline
   60 & 0.95258 & 0.95260 & 0.95258$\pm$0.68E-04 \\  \hline
   70 & 0.94938 & 0.94941 & 0.94939$\pm$0.71E-04 \\  \hline
   80 & 0.94653 & 0.94656 & 0.94654$\pm$0.73E-04 \\  \hline
   90 & 0.94395 & 0.94399 & 0.94397$\pm$0.75E-04 \\  \hline
  100 & 0.94161 & 0.94165 & 0.94162$\pm$0.76E-04 \\  \hline
\end{tabular}
\caption{  Numerical results on the compensation for 
the $Q^2$-independent contributions.
 The second figures show
$D^f(n)^{-1}$, while the third ones are approximated contributions by
 $D^{f,A}(n)$. Results by the shower algorithm are given 
in the last column      }
\end{table}
\begin{table}[h]
\begin{tabular}{|c|c|c|c|c|} \hline
mom &  LL order(analyt.)   & NLL order(analyt.) 
& $x_{b1}(1-t_1)x_{b2}(1-t_2)$ & $q^2/S$ \\ \hline
    2 & 0.92649& 0.92677 & 0.92655$\pm$0.37E-04 & 0.92554$\pm$0.38E-04 \\  \hline
    3 & 0.88754& 0.88799 & 0.88789$\pm$0.51E-04 & 0.88723$\pm$0.51E-04 \\  \hline
    4 & 0.86085& 0.86143 & 0.86140$\pm$0.57E-04 & 0.86090$\pm$0.57E-04 \\  \hline
    5 & 0.84055& 0.84124 & 0.84124$\pm$0.61E-04 & 0.84085$\pm$0.61E-04 \\  \hline
    6 & 0.82421& 0.82498 & 0.82500$\pm$0.63E-04 & 0.82467$\pm$0.63E-04 \\  \hline
    7 & 0.81056& 0.81141 & 0.81143$\pm$0.65E-04 & 0.81115$\pm$0.65E-04 \\  \hline
    8 & 0.79886& 0.79977 & 0.79979$\pm$0.67E-04 & 0.79955$\pm$0.67E-04 \\  \hline
    9 & 0.78863& 0.78960 & 0.78963$\pm$0.68E-04 & 0.78941$\pm$0.68E-04 \\  \hline
   10 & 0.77956& 0.78058 & 0.78060$\pm$0.69E-04 & 0.78041$\pm$0.69E-04 \\  \hline
   11 & 0.77142& 0.77249 & 0.77251$\pm$0.70E-04 & 0.77233$\pm$0.70E-04 \\  \hline
   12 & 0.76404& 0.76516 & 0.76517$\pm$0.71E-04 & 0.76501$\pm$0.71E-04 \\  \hline
   13 & 0.75730& 0.75846 & 0.75847$\pm$0.72E-04 & 0.75832$\pm$0.72E-04 \\  \hline
   14 & 0.75110& 0.75229 & 0.75230$\pm$0.73E-04 & 0.75217$\pm$0.73E-04 \\  \hline
   15 & 0.74536& 0.74659 & 0.74660$\pm$0.74E-04 & 0.74647$\pm$0.74E-04 \\  \hline
   16 & 0.74002& 0.74128 & 0.74129$\pm$0.74E-04 & 0.74117$\pm$0.75E-04 \\  \hline
   17 & 0.73504& 0.73633 & 0.73633$\pm$0.75E-04 & 0.73622$\pm$0.75E-04 \\  \hline
   18 & 0.73036& 0.73168 & 0.73168$\pm$0.76E-04 & 0.73158$\pm$0.76E-04 \\  \hline
   19 & 0.72596& 0.72731 & 0.72730$\pm$0.77E-04 & 0.72721$\pm$0.77E-04 \\  \hline
   20 & 0.72180& 0.72318 & 0.72317$\pm$0.77E-04 & 0.72308$\pm$0.77E-04 \\  \hline
   30 & 0.68965& 0.69123 & 0.69120$\pm$0.81E-04 & 0.69114$\pm$0.81E-04 \\  \hline
   40 & 0.66759& 0.66932 & 0.66928$\pm$0.83E-04 & 0.66924$\pm$0.83E-04 \\  \hline
   50 & 0.65093& 0.65277 & 0.65271$\pm$0.83E-04 & 0.65268$\pm$0.83E-04 \\  \hline
   60 & 0.63759& 0.63952 & 0.63946$\pm$0.84E-04 & 0.63943$\pm$0.84E-04 \\  \hline
   70 & 0.62652& 0.62853 & 0.62846$\pm$0.84E-04 & 0.62843$\pm$0.84E-04 \\  \hline
   80 & 0.61707& 0.61914 & 0.61907$\pm$0.85E-04 & 0.61905$\pm$0.85E-04 \\  \hline
   90 & 0.60885& 0.61098 & 0.61090$\pm$0.85E-04 & 0.61089$\pm$0.85E-04 \\  \hline
  100 & 0.60158& 0.60377 & 0.60369$\pm$0.86E-04 & 0.60367$\pm$0.86E-04 \\  \hline
\end{tabular}
\caption{  Numerical results in the NLL order. 
 The second figures show moments in the LL order,
 while the third ones are the NLL order results. Both are calculated
analytically.
 Results  on  $x_{b1}(1-t_1)x_{b2}(1-t_2)$ and
$q^2/S$  obtained by our generator  are given
in the last two columns.
 $S=10^4 \GeV^2$,  $\mu^2 =0.25\times 10^{-6}\GeV^2 $.      }
\end{table}
\begin{table}[h]
\begin{tabular}{|c|c|c|c|c|} \hline
mom &  LL order(analyt.)   & NLL order(analyt.) 
& $x_{b1}(1-t_1)x_{b2}(1-t_2)$ & $q^2/S$ \\ \hline
    2 & 0.91309& 0.91341 & 0.91318 $\pm$0.53E-04 & 0.91216 $\pm$0.55E-04 \\  \hline
    3 & 0.86757& 0.86809 & 0.86795 $\pm$0.66E-04 & 0.86729 $\pm$0.67E-04 \\  \hline
    4 & 0.83658& 0.83726 & 0.83717 $\pm$0.74E-04 & 0.83668 $\pm$0.74E-04 \\  \hline
    5 & 0.81315& 0.81394 & 0.81387 $\pm$0.79E-04 & 0.81348 $\pm$0.80E-04 \\  \hline
    6 & 0.79436& 0.79525 & 0.79518 $\pm$0.84E-04 & 0.79486 $\pm$0.84E-04 \\  \hline
    7 & 0.77872& 0.77969 & 0.77962 $\pm$0.87E-04 & 0.77934 $\pm$0.87E-04 \\  \hline
    8 & 0.76535& 0.76639 & 0.76632 $\pm$0.90E-04 & 0.76608 $\pm$0.90E-04 \\  \hline
    9 & 0.75370& 0.75481 & 0.75472 $\pm$0.93E-04 & 0.75451 $\pm$0.93E-04 \\  \hline
   10 & 0.74338& 0.74455 & 0.74446 $\pm$0.95E-04 & 0.74427 $\pm$0.95E-04 \\  \hline
   11 & 0.73415& 0.73537 & 0.73527 $\pm$0.97E-04 & 0.73510 $\pm$0.97E-04 \\  \hline
   12 & 0.72579& 0.72706 & 0.72696 $\pm$0.98E-04 & 0.72681 $\pm$0.98E-04 \\  \hline
   13 & 0.71818& 0.71949 & 0.71938 $\pm$0.10E-03 & 0.71924 $\pm$0.10E-03 \\  \hline
   14 & 0.71118& 0.71253 & 0.71242 $\pm$0.10E-03 & 0.71229 $\pm$0.10E-03 \\  \hline
   15 & 0.70471& 0.70610 & 0.70598 $\pm$0.10E-03 & 0.70586 $\pm$0.10E-03 \\  \hline
   16 & 0.69871& 0.70013 & 0.70001 $\pm$0.10E-03 & 0.69990 $\pm$0.10E-03 \\  \hline
   17 & 0.69311& 0.69456 & 0.69444 $\pm$0.10E-03 & 0.69433 $\pm$0.10E-03 \\  \hline
   18 & 0.68786& 0.68935 & 0.68922 $\pm$0.11E-03 & 0.68912 $\pm$0.11E-03 \\  \hline
   19 & 0.68293& 0.68444 & 0.68431 $\pm$0.11E-03 & 0.68422 $\pm$0.11E-03 \\  \hline
   20 & 0.67827& 0.67982 & 0.67968 $\pm$0.11E-03 & 0.67959 $\pm$0.11E-03 \\  \hline
   30 & 0.64245& 0.64421 & 0.64405 $\pm$0.11E-03 & 0.64399 $\pm$0.11E-03 \\  \hline
   40 & 0.61806& 0.61997 & 0.61979 $\pm$0.11E-03 & 0.61975 $\pm$0.11E-03 \\  \hline
   50 & 0.59973& 0.60175 & 0.60156 $\pm$0.11E-03 & 0.60153 $\pm$0.11E-03 \\  \hline
   60 & 0.58513& 0.58724 & 0.58705 $\pm$0.11E-03 & 0.58702 $\pm$0.11E-03 \\  \hline
   70 & 0.57304& 0.57524 & 0.57503 $\pm$0.11E-03 & 0.57501 $\pm$0.11E-03 \\  \hline
   80 & 0.56277& 0.56503 & 0.56482 $\pm$0.11E-03 & 0.56480 $\pm$0.11E-03 \\  \hline
   90 & 0.55385& 0.55616 & 0.55595 $\pm$0.11E-03 & 0.55594 $\pm$0.11E-03 \\  \hline
  100 & 0.54599& 0.54835 & 0.54814 $\pm$0.11E-03 & 0.54812 $\pm$0.11E-03 \\  \hline
\end{tabular}
\caption{  
 The same as Table 5 except $S$. Here
 $S=10^6 \GeV^2$.      }
\end{table}
\begin{table}[h]
\begin{tabular}{||c|c|c||c|c||c|c||} \hline
 & \multicolumn{2}{c|}{$S=10^4\GeV^2$} &
   \multicolumn{2}{c|}{$S=10^6\GeV^2$} &
   \multicolumn{2}{c|}{$S=10^8\GeV^2$} \\ \hline
mom &generator     & shower  
 &generator    & shower   
 &generator     & shower  
 \\ \hline
    2  &-0.13218\% &-0.02363\% &-0.13773\% &-0.02573\% &-0.13908\% &-0.02333\%  \\  \hline
    3  &-0.08570\% &-0.01171\% &-0.09262\% &-0.01624\% &-0.09109\% &-0.01249\%  \\  \hline
    4  &-0.06095\% &-0.00395\% &-0.06951\% &-0.01075\% &-0.06612\% &-0.00565\%  \\  \hline
    5  &-0.04660\% &-0.00012\% &-0.05688\% &-0.00872\% &-0.05194\% &-0.00254\%  \\  \hline
    6  &-0.03770\% & 0.00182\% &-0.04917\% &-0.00843\% &-0.04318\% &-0.00130\%  \\  \hline
    7  &-0.03167\% & 0.00271\% &-0.04438\% &-0.00898\% &-0.03751\% &-0.00107\%  \\  \hline
    8  &-0.02738\% & 0.00300\% &-0.04110\% &-0.00979\% &-0.03350\% &-0.00136\%  \\  \hline
    9  &-0.02432\% & 0.00291\% &-0.03882\% &-0.01073\% &-0.03063\% &-0.00180\%  \\  \hline
   10  &-0.02203\% & 0.00269\% &-0.03720\% &-0.01182\% &-0.02859\% &-0.00239\%  \\  \hline
   11  &-0.02032\% & 0.00233\% &-0.03617\% &-0.01278\% &-0.02700\% &-0.00300\%  \\  \hline
   12  &-0.01895\% & 0.00196\% &-0.03521\% &-0.01375\% &-0.02562\% &-0.00347\%  \\  \hline
   13  &-0.01767\% & 0.00158\% &-0.03461\% &-0.01473\% &-0.02447\% &-0.00396\%  \\  \hline
   14  &-0.01675\% & 0.00133\% &-0.03410\% &-0.01558\% &-0.02386\% &-0.00474\%  \\  \hline
   15  &-0.01607\% & 0.00094\% &-0.03385\% &-0.01643\% &-0.02307\% &-0.00524\%  \\  \hline
   16  &-0.01538\% & 0.00054\% &-0.03342\% &-0.01714\% &-0.02269\% &-0.00590\%  \\  \hline
   17  &-0.01467\% & 0.00027\% &-0.03326\% &-0.01785\% &-0.02229\% &-0.00641\%  \\  \hline
   18  &-0.01435\% &-0.00014\% &-0.03307\% &-0.01857\% &-0.02187\% &-0.00678\%  \\  \hline
   19  &-0.01389\% &-0.00041\% &-0.03302\% &-0.01914\% &-0.02159\% &-0.00730\%  \\  \hline
   20  &-0.01369\% &-0.00083\% &-0.03310\% &-0.01986\% &-0.02160\% &-0.00798\%  \\  \hline
   30  &-0.01288\% &-0.00420\% &-0.03415\% &-0.02530\% &-0.02216\% &-0.01300\%  \\  \hline
   40  &-0.01300\% &-0.00642\% &-0.03549\% &-0.02871\% &-0.02369\% &-0.01672\%  \\  \hline
   50  &-0.01348\% &-0.00827\% &-0.03673\% &-0.03141\% &-0.02525\% &-0.01966\%  \\  \hline
   60  &-0.01407\% &-0.00969\% &-0.03814\% &-0.03355\% &-0.02690\% &-0.02226\%  \\  \hline
   70  &-0.01448\% &-0.01082\% &-0.03911\% &-0.03529\% &-0.02850\% &-0.02451\%  \\  \hline
   80  &-0.01486\% &-0.01163\% &-0.04018\% &-0.03681\% &-0.02968\% &-0.02619\%  \\  \hline
   90  &-0.01522\% &-0.01228\% &-0.04100\% &-0.03794\% &-0.03082\% &-0.02766\%  \\  \hline
  100  &-0.01557\% &-0.01292\% &-0.04194\% &-0.03921\% &-0.03194\% &-0.02892\%  \\  \hline
\end{tabular}
\caption{  Differences between the analytic results and the generator's or
the shower's ones.
 The ratios of them are shown.
 The columns with $q^2/S$ denote  the ratios by the generator.
 We neglect the statistical  error. 
 Here the shower and the generator  compensate for the $Q^2$-independent
contributions.
     }
\end{table}
\begin{table}[h]
\begin{tabular}{|c|c|c|} \hline
mom &    Eq.(76)/$(\alpha_0/2\pi)^2 $  & Eq.(76)
 \\ \hline
    1 &  3.3366 & 0.4503E-05 \\  \hline
    2 &  4.4792 & 0.6045E-05 \\  \hline
    3 &  5.0860 & 0.6864E-05 \\  \hline
    4 &  5.4672 & 0.7378E-05 \\  \hline
    5 &  5.7301 & 0.7733E-05 \\  \hline
    6 &  5.9227 & 0.7993E-05 \\  \hline
    7 &  6.0702 & 0.8192E-05 \\  \hline
    8 &  6.1868 & 0.8350E-05 \\  \hline
    9 &  6.2813 & 0.8477E-05 \\  \hline
   10 &  6.3594 & 0.8583E-05 \\  \hline
   11 &  6.4252 & 0.8671E-05 \\  \hline
   12 &  6.4812 & 0.8747E-05 \\  \hline
   13 &  6.5296 & 0.8812E-05 \\  \hline
   14 &  6.5718 & 0.8869E-05 \\  \hline
   15 &  6.6089 & 0.8919E-05 \\  \hline
   16 &  6.6418 & 0.8964E-05 \\  \hline
   17 &  6.6712 & 0.9003E-05 \\  \hline
   18 &  6.6976 & 0.9039E-05 \\  \hline
   19 &  6.7214 & 0.9071E-05 \\  \hline
   20 &  6.7430 & 0.9100E-05 \\  \hline
   30 &  6.8840 & 0.9291E-05 \\  \hline
   40 &  6.9574 & 0.9390E-05 \\  \hline
   50 &  7.0023 & 0.9450E-05 \\  \hline
   60 &  7.0327 & 0.9491E-05 \\  \hline
   70 &  7.0546 & 0.9521E-05 \\  \hline
   80 &  7.0711 & 0.9543E-05 \\  \hline
   90 &  7.0840 & 0.9560E-05 \\  \hline
  100 &  7.0944 & 0.9574E-05 \\  \hline
\end{tabular}
\caption{  Numerical results on $Diff$ of Eq.(76) in Appendix C.
   These differences
 are shown in the third column, while values in the second one are
calculated by dividing them by $(\alpha_0/2\pi)^2$.    }
\end{table}
\begin{figure}[b]
\centering
\epsfxsize=0.65\textwidth
\epsfbox{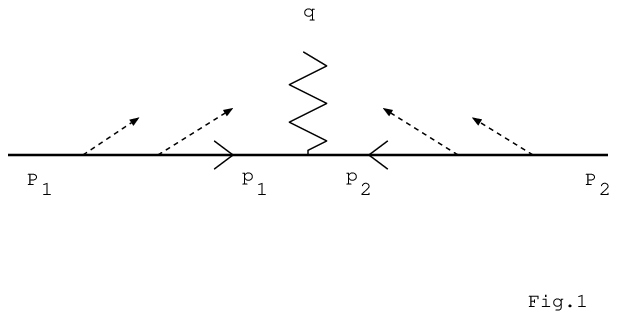}
\caption{ The schematics of the annihilation process.
 The  electron(positron) has the momentum $P_1$($P_2$) initially,
and does $p_1$($p_2$) after the branching process.
The momentum of the virtual photon  is denoted by $q$. }
\end{figure}

\end{document}